\documentclass[10pt, journal, letterpaper]{IEEEtran}
\IEEEoverridecommandlockouts
 \usepackage{todonotes}
\usepackage{cite}
\usepackage{amsmath,amssymb,amsfonts}
\usepackage{algorithmic}
\usepackage{algorithm}
\usepackage{tabularray}
\usepackage{changes}
\makeatletter
\newcommand{\multiline}[1]{%
  \begin{tabularx}{\dimexpr\linewidth-\ALG@thistlm}[t]{@{}X@{}}
    #1
  \end{tabularx}
}
\makeatother
\usepackage{graphicx}
\usepackage{textcomp}
\usepackage{xcolor}
\usepackage{tabularx}
\usepackage{tikz}
\usepackage{pgfplots}
\usepackage{siunitx}
\usepackage{bm}
\usepackage{bbm}
\usepackage{setspace}
\usepackage[caption=false, font=footnotesize]{subfig}
\usepackage[normalem]{ulem}
\newcolumntype{Y}{>{\small\raggedright\arraybackslash}X}
\def\BibTeX{{\rm B\kern-.05em{\sc i\kern-.025em b}\kern-.08em
    T\kern-.1667em\lower.7ex\hbox{E}\kern-.125emX}}

\usepackage{amsthm}
\theoremstyle{plain}
\newtheorem{theorem}{Theorem}
\newtheorem{definition}{Definition}
\pgfplotsset{compat=1.18}
\usepackage{subfig}
\usepackage{setspace}

\begin{document}
\newlength\fwidth
\newlength\fheight
\renewcommand{\algorithmiccomment}[1]{\hfill{$\triangleright$\ #1}}
\title{Decentralized Online Learning in Task Assignment Games for Mobile Crowdsensing}


\author{
    \IEEEauthorblockN{\normalsize Bernd Simon\thanks{This work has been funded by the German Research Foundation (DFG) 
within the Collaborative Research Center (CRC) 1053 MAKI and has been supported by the BMBF project Open6GHub (Nr. 16KISK014). This research was also, in part, supported by the U.S. National Science Foundation under Grant ECR-EDU-2201641.}\IEEEauthorrefmark{1}, Andrea Ortiz\IEEEauthorrefmark{1}, Walid Saad\IEEEauthorrefmark{2} and Anja Klein\IEEEauthorrefmark{1}}\\
    \IEEEauthorblockA{\small \IEEEauthorrefmark{1}Communication Engineering Lab, Technische Universit\"{a}t Darmstadt, Darmstadt, Germany.
    \\ \IEEEauthorrefmark{2}Wireless@VT, Bradley Department of Electrical and Computer Engineering, Virginia Tech, Arlington, VA, USA. \\
    Emails: \{b.simon, a.ortiz,  a.klein\}@nt.tu-darmstadt.de}, walids@vt.edu.}

\newcommand{\new} [1]{{\color{blue}#1}}
\newcommand{\mv} [2]{{\color{green} \textbf{Was in Section \ref{#1};} #2}}
\newcommand{\out} [1]{{\color{red} \sout{#1}}}
\newcommand{\algo} {\textit{Collision-Avoidance Multi-Armed Bandit with Strategic Free Sensing}}
\newcommand{\algoShort} {\textit{CA-MAB-SFS}}
\newcommand{\mb}[1]{\mathbf{#1}}
\maketitle
\newcommand{\timeindex}{t}
\newcommand{\timehorizon}{T}

\newcommand{\MUindex}{k}
\newcommand{\setOfMUs}{\mathcal{K}}
\newcommand{\numberOfMUs}{K}
\newcommand{\MUwithIndex}[1]{\text{MU}~#1}
\newcommand{\MUk}{\MUwithIndex{\MUindex}}
\newcommand{\energySensitivity}{{\color{red} REMOVE ME}}
\newcommand{\effortMargin}{\beta}
\newcommand{\localFrequency}{f_k^{\mathrm{local}}}

\newcommand{\numberOfTasks}{N}
\newcommand{\setOfTaskTypes}{\mathcal{Z}}
\newcommand{\numberOfTaskTypes}{Z}
\newcommand{\taskTypeIndex}{z}
\newcommand{\taskTypeZ}{\taskTypeIndex}
\newcommand{\taskIndex}{n}
\newcommand{\setOfTasks}{\mathcal{A}}
\newcommand{\taskWithIndex}[1]{{a}_{#1}}
\newcommand{\taskTypeWithIndex}[1]{\mathrm{a}_{#1}}
\newcommand{\taskn}{\taskTypeWithIndex{\taskIndex}}
\newcommand{\deadlineOfTask}{\tau_{\taskTypeIndex}^{\mathrm{max}}}
\newcommand{\taskSize}{s_\taskTypeIndex}

\newcommand{\setOfTasksWithType}[1]{\setOfTasks_{{#1},\timeindex}}
\newcommand{\setOfTasksWithTypeZ}{\setOfTasksWithType{\taskTypeIndex}}
\newcommand{\mappingFunction}{g_\timeindex}

\newcommand{\MUTaskAndTimeIndex}{_{\MUindex,\taskIndex,\timeindex}}
\newcommand{\MUTaskIndex}{_{\MUindex,\taskIndex}}

\newcommand{\elementOfAssignmentMatrixWithIndex}[1]{x_{#1}}
\newcommand{\elementOfAssignmentMatrix}{\elementOfAssignmentMatrixWithIndex{\MUTaskAndTimeIndex}}
\newcommand{\assignmentMatrix}{\boldsymbol{X}_{\timeindex}}

\newcommand{\paymentFunction}{P^{\mathrm{effort}}}
\newcommand{\costFunction}{C_k^{\mathrm{effort}}}

\newcommand{\taskComplexity}{c_{z}}
\newcommand{\sensingTime}{\tau^{\mathrm{sense}}\MUTaskAndTimeIndex}
\newcommand{\expectedSensingTime}{\bar{\tau}^{\mathrm{sense}}_{\MUindex,\taskTypeIndex}}
\newcommand{\communicationTime}{\tau^{\mathrm{comm}}\MUTaskAndTimeIndex}
\newcommand{\expectedCommunicationTime}{\bar{\tau}^{\mathrm{comm}}_{\MUindex,\taskTypeIndex}}
\newcommand{\computationTime}{\tau^{\mathrm{comp}}\MUTaskAndTimeIndex}
\newcommand{\totalEnergy}{E\MUTaskAndTimeIndex}
\newcommand{\totalTime}{\tau\MUTaskAndTimeIndex}
\newcommand{\expectedUtilityTotalTimeWithIndex}[1]{\bar{\tau}^{\mathrm{MU}}_{#1}}
\newcommand{\expectedTotalTime}{\expectedUtilityMUWithIndex{\MUindex,\taskIndex}}

\newcommand{\txPower}{p_k^\mathrm{comm}}
\newcommand{\compPower}{p_k^\mathrm{comp}}

\newcommand{\payment}{P\MUTaskAndTimeIndex}
\newcommand{\MUpaymentProposal}{\hat{P}_{\MUindex,\taskTypeIndex}}

\newcommand{\utilityMU}{U^{\mathrm{MU}}_{\MUindex,\taskIndex,\timeindex}}
\newcommand{\expectedUtilityMUWithIndex}[1]{\bar{U}^{\mathrm{MU}}_{#1}}
\newcommand{\expectedUtilityMU}{\expectedUtilityMUWithIndex{\MUindex,\taskTypeIndex}}

\newcommand{\utilityTask}{U^{\mathrm{MCSP}}_{\MUindex,\taskIndex,\timeindex}}
\newcommand{\expectedUtilityTaskWithIndex}[1]{\bar{U}^{\mathrm{MCSP}}_{#1}}
\newcommand{\expectedUtilityTask}{\expectedUtilityTaskWithIndex{\MUindex,\taskTypeIndex}}

\newcommand{\rewardTaskCompletion}{w_{\taskTypeIndex,\timeindex}}
\newcommand{\MUpreferenceWithIndex}[1]{\succeq^{\mathrm{MU}}_{#1}}
\newcommand{\MUpreference}{\MUpreferenceWithIndex{\MUindex}}
\newcommand{\MUstrictPreferenceWithIndex}[1]{\succ^{\mathrm{MU}}_{#1}}
\newcommand{\TaskpreferenceWithIndex}[1]{\succeq^{\mathrm{MCSP}}_{#1}}
\newcommand{\Taskpreference}{\TaskpreferenceWithIndex{\taskTypeIndex}}

\newcommand{\socialWelfare}{U^{\mathrm{SW}}_{\timeindex}(\assignmentMatrix)}

\newcommand{\stableTaskForMUk}{\taskWithIndex{k}^{\mathrm{stable}}}
\newcommand{\stableExpectedUtilityMU}{\bar{U}^{\mathrm{MU,stable}}_{\MUindex}}

\newcommand{\estimatedUtilityWithIndex}[1]{\hat{U}_{#1}}

\newcommand{\sensingOfferWithIndex}[1]{O_{#1}}
\newcommand{\sensingOffer}{\sensingOfferWithIndex{\MUindex,\timeindex}}
\newcommand{\sensingAccepted}{\bar{O}_{\MUindex,\timeindex}}

\newcommand{\freeSensingAdjustmentPara}{\epsilon^{\mathrm{a}}}
\newcommand{\freeSensingStopPara}{\epsilon^{\mathrm{e}}}
\newcommand{\freeSensingSensitivityPara}{\epsilon^{\mathrm{s}}}

\newcommand{\umax}{\Delta\mathrm{U}}
\begin{abstract}
The problem of coordinated data collection is studied for a mobile crowdsensing (MCS) system. A mobile crowdsensing platform (MCSP) sequentially publishes sensing tasks to the available mobile units (MUs) that signal their willingness to participate in a task by sending sensing offers back to the MCSP. From the received offers, the MCSP decides the task assignment. A stable task assignment must address two challenges: the MCSP’s and MUs’ conflicting goals, and the uncertainty about the MUs' required efforts and preferences. To overcome these challenges a novel decentralized approach combining matching theory and online learning, called collision-avoidance multi-armed bandit with strategic free sensing (CA-MAB-SFS), is proposed. The task assignment problem is modeled as a matching game considering the MCSP's and MUs' individual goals while the MUs learn their efforts online. Our innovative “free-sensing” mechanism significantly improves the MU’s learning process while reducing collisions during task allocation. The stable regret of CA-MAB-SFS, i.e., the loss of learning, is analytically shown to be bounded by a sublinear function, ensuring the convergence to a stable optimal solution. Simulation results show that CA-MAB-SFS increases the MUs' and the MCSP's satisfaction compared to state-of-the-art methods while reducing the average task completion time by at least $16\,\%$.

\end{abstract}

\vspace{-0.4cm}
\section{Introduction}
\label{sec:Introduction}

Mobile devices such as smartphones and wearables are ubiquitous. 
In fact, by 2025 the number of mobile devices in the world is expected to reach 18.2 billion~\cite{Statista2021}. 
As these mobile devices are usually equipped with different sensors, they can be leveraged to collectively perform sensing tasks via mobile crowdsensing (MCS) techniques, e.g., see~\cite{Capponi2018crowdSensingSurvey} and~\cite{Nie2019_Incentive}.
In MCS, a group or ``crowd'' of mobile units (MUs) performs sensing tasks.
Compared with conventional wireless sensor networks, MCS has much lower infrastructure costs, higher coverage, and a wider range of applications due to the mobility of the MUs \cite{Ma2014, Gong2018crowdsensingSurvey, Dongare2022}.
It is, therefore, no surprise that the interest in MCS has steadily increased across academia and industry.

A typical MCS system is composed of one or multiple data requesters, an MCS platform (MCSP), and multiple MUs \cite{Dongare2022} and \cite{Wang2019}.
The data requesters submit their sensing requests to the MCSP who acts as the intermediary between the data requesters and the MUs.
Particularly, the MCSP converts the sensing requests into sensing tasks, and publishes the tasks to the MUs including information about their type. 
The MUs independently decide whether to participate or not in each published task.
This decision is selfishly and individually made by each MU depending on the effort needed to perform the task and the expected payment from the MCSP~\cite{ahuja2019dynamic}. 
The MUs signal their willingness to participate in a task by sending a sensing offer to the MCSP containing a payment proposal, i.e., the number of monetary units the MU is charging the MCSP for performing the task.
Based on the offers of the MUs, the MCSP then decides which task is assigned to each MU by sending them an acknowledgment to their sensing proposal.
The revenue of the MCSP depends on its own earnings, i.e, the net payments received from the data requesters for their service after paying the MUs for performing the sensing tasks.
The MUs' satisfaction depends on the number of sensing offers that were accepted by the MSCP. 


\vspace*{-0.4cm}
\subsection{Research Challenges}
\label{sec:challenges}
The assignment of the sensing tasks to the requesting MUs is a fundamental problem that will be a key determinant of the success of MCS. 
This assignment must be able to maximize both the satisfaction of the MUs, and the MCSP revenues~\cite{Wang2018_opportunities}, such that the MCSP and MUs do not have any incentive to deviate from the chosen task assignment. 
To achieve this the MCS must overcome two major challenges, as discussed next.
\subsubsection{Considering multiple utility functions} 
The first key challenge is that the interests of the MCSP and the MUs are not aligned.
Each participant in MCS, including MUs and the MCSP, have their own utility functions with technical and economic components.
The MUs want to maximize the payment obtained from the MCSP while minimizing the expounded effort, in terms of energy consumption and completion time.
The MCSP maximizes its revenue by assigning tasks to MUs which require a lower payment. 
Consequently, the MCSP and the MUs may act selfishly to maximize their own revenues.
\subsubsection{Incomplete information}
The second key challenge is that the MUs and the MCSP do not have complete information about the MCS system.
This incomplete information spans two components: 1) incomplete information about the tasks and 2) incomplete information about the other participants.
Firstly, the effort that an MU must spend to execute a given task is often not known beforehand.
For instance, the MUs know the task types from the list of published tasks, but they have to explore how much effort is required to complete the tasks.
Moreover, the characteristics of the published tasks and the MU's conditions, such as the communication rate, change over time depending on factors like the sensing preferences of the data requesters and the mobility of the MUs. 
Both the task characteristics and the MU's conditions, are therefore appropriately modeled as random processes whose probability distributions are not known a priori.
Furthermore, the MCSP does not know the effort that the MUs need to complete the sensing tasks, and the MUs can only measure this effort by executing that particular task.

Secondly, the MUs do not know what task types the other MUs prefer.
This may result in colliding sensing offers and unstable assignments.
A collision occurs when more than the allowable number of MUs send sensing offers for the same task type.
Such concurrent sensing offers occur because the MUs cannot observe each other's sensing offers.
Therefore, they are unaware of the effort required by other MUs to perform a task.
Collisions should be avoided because they lead to performance degradation as the sensing capabilities of the MUs involved in the collision cannot be used until the next task arrives.
In practical MCS systems, these two key research challenges have to be jointly solved because they incorporate the main characteristics of the MCSP and the MUs.

\vspace*{-0.4cm}
\subsection{Related Works}
Prior works~\cite{Nie2019_Incentive,Wang2019,Dongare2022} and \cite{Gong2015Utility,Karaliopoulos2015Opportunistic,Yucel2022,Simon2022, Wang2019_Auction,Xiao2017, Wang2018, Zhang2021, Gao2022, Xiao2021} that attempted to address the aforementioned challenges related to MCS task assignment can be categorized into three directions: 
i) Optimization approaches, such as in \cite{Gong2015Utility} and~\cite{Karaliopoulos2015Opportunistic}, 
ii) Game theory approaches, such as in \cite{Nie2019_Incentive, Wang2019} and \cite{Yucel2022, Simon2022, Wang2019_Auction}, and 
iii) Online learning approaches, such as in \cite{Dongare2022} and \cite{Xiao2017, Wang2018, Zhang2021, Gao2022, Xiao2021}.
Although the authors in \cite{Gong2015Utility} and \cite{Karaliopoulos2015Opportunistic} find optimal allocation policies that maximize the MCSP's utility, the MU's utility functions are not considered. We argue that this limitation to a single utility function is not realistic. Moreover, it requires complete non-causal information about the MCS system. 

Following a game theory approach, the authors in  \cite{Nie2019_Incentive} investigate an optimal incentive mechanism for the MCS using a two-stage Stackelberg game. Their goal is to efficiently recruit MUs to perform the available sensing tasks while assuming payments to be fixed in advance.
In\cite{Wang2019}, the MU's effort is assumed to depend on its location.
The authors propose a privacy-preserving approach to obtain information about the MU's location and thus, estimate the MU's efforts.
The authors in \cite{Yucel2022} use matching theory to balance the preferences of the MCSP and the MUs while assuming the payments by the MCSP are fixed in advance.
Similarly, assuming known preferences for the MCSP and the MUs, the authors in \cite{Simon2022} formulate a two-stage matching problem to maximize the coverage in a MCS system. 
Following a social welfare maximization approach, the authors in~\cite{Wang2019_Auction} propose an auction-based method to balance the MCSP and MU's interests when assigning the sensing tasks. The use of these game-theory-based approaches allows the consideration of the conflicting goals of the MCSP and MUs. However, similar to the optimization approaches~\cite{Gong2015Utility} and \cite{Karaliopoulos2015Opportunistic},  the game theory approaches~\cite{Nie2019_Incentive, Wang2019} and \cite{Yucel2022, Simon2022, Wang2019_Auction} are subjected to the strict requirement that information about the MUs' costs and/or payment requests is known in advance. 
This requirement makes these approaches infeasible in practical systems, as the tasks' characteristics and the effort to complete tasks are not known a priori and may change over time.

The problem of task assignment under unknown MU efforts is investigated in \cite{Xiao2017, Wang2018, Zhang2021, Gao2022, Xiao2021}.     
In \cite{Xiao2017}, the authors propose a location-prediction-based online task assignment strategy in which the MU's effort depends on its location in a mobile social network. 
In \cite{Wang2018}, Lyapunov optimization is used to derive a task assignment policy that maximizes the gain of the MCSP. 
The authors in \cite{Zhang2021} propose prediction methods to estimate the MU's effort at the MCSP.
The task pricing problem in a point-to-point MCS system is considered in \cite{Gao2022}, where a two-stage mean field approximation Stackelberg differential game is used to model the MCSP-MU interaction.
Combinatorial multi-armed bandits are considered in \cite{Xiao2021} to maximize the expected quality of the data received at the MSCP.
Even though the solutions in~\cite{Xiao2017, Wang2018, Zhang2021, Gao2022, Xiao2021} overcome the requirement of complete non-causal information about the MCS, they are limited to a single utility function, i.e., they only consider either the MCSP's or the MUs' perspective when the MU's efforts are unknown.

Clearly, as discussed, the prior art is limited in several ways. 
The conflicting interests of the MCSP and MUs under realistic conditions, i.e., when the MU's efforts are not known in advance, have not been considered yet.
Furthermore, the prior art does not consider the problem of collision in the online learning scenario.
Collisions may significantly reduce the overall performance and therefore need to be avoided.
This open problem of online learning for the task assignment can be cast as a multi-player multi-armed bandit problem \cite{Magesh2022multiplayerMAB}.
In the learning literature, multi-player multi-armed bandits  have been investigated under some simplifying assumptions. 
For example, assuming that there are no individual preferences, the authors in~\cite{Shu2021multiplayer} propose to divide the reward among  colliding agents to improve the learning speed. 
In~\cite{liu2021bandit}, a multi-armed bandit with a collision-avoidance mechanism is proposed. The authors assume that there are no individual costs or payments associated to the decisions in order to allow each player to learn its own preferences while avoiding collisions with competitors. 
Centralized and decentralized learning strategies are compared in~\cite{liu2020competingBandits}, where the effect of sharing the learned preferences is analyzed.
This work assumes a cooperative setting, in which all agents communicate their decisions with all other agents.
Despite considering multi-agent multi-armed bandits, the solutions in~\cite{Magesh2022multiplayerMAB,Shu2021multiplayer,liu2021bandit,liu2020competingBandits} cannot be applied to the task allocation problem in MCS. Their simplifying assumptions clash with the requirements of MCS. Specifically, the MCSP and the MUs have individual preferences according to their capabilities and conditions. Moreover, the allocation of task implies an effort for the MUs and a payment for the MCSP, and the strict privacy constraints and communication overhead requirements limit the communication between the agents.

\begin{figure*}[t]
 \centering
  \subfloat[The MCSP broadcasts the list of available tasks to all MUs.\label{fig:system_model_a}]{%
       \includegraphics[width=0.23\linewidth]{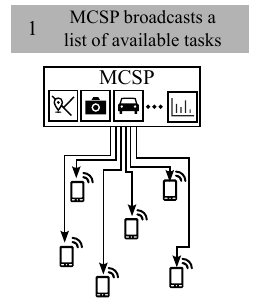}}
    \hfill
  \subfloat[The MUs send a sensing proposal to the MCSP. The red circle represents a collision.\label{fig:system_model_b}]{%
        \includegraphics[width=0.23\linewidth]{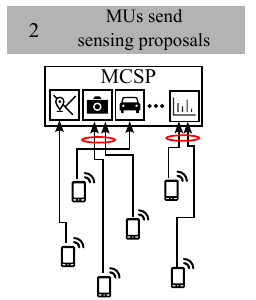}}
    \hfill
  \subfloat[The MCSP accepts or rejects the sensing proposals of the MUs.\label{fig:system_model_c}]{%
        \includegraphics[width=0.23\linewidth]{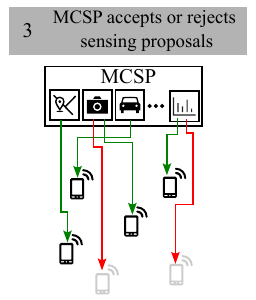}}
    \hfill
  \subfloat[The MUs perform the task, transmit the result, and receive their payment.\label{fig:system_model_d}]{%
        \includegraphics[width=0.23\linewidth]{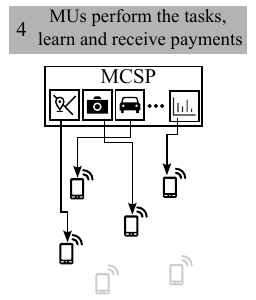}}
  \caption{Overview of the system model.}
  \label{fig:system_model} 
  \vspace{-0.55cm}
\end{figure*}

\vspace{-0.2cm}
\subsection{Contributions}
The main contribution of this paper is a novel decentralized task assignment scheme for MCS that can improve the satisfaction of the MUs and the MCSP, which are considered to be individual rational decision makers with \emph{incomplete information}.
In the studied MCS system, the effort required for each task in terms of completion time and energy consumption is not known initially, which leads to a difficult learning problem.
Using existing online learning solutions leads to many collisions between the MUs, which results in a high overhead and degraded overall system performance.
In particular, we propose a novel decentralized algorithm termed collision-avoidance multi-armed bandit with strategic free sensing (\algoShort{}), whose goal is to find a stable task assignment, i.e., a task assignment where neither the MUs nor the MCSP have an incentive to change the task assignment.
Our contributions can therefore be summarized as follows:
\begin{itemize}
    \item To balance the conflicting interests of the MCSP and the MUs, we propose the use of a novel decentralized online learning strategy which leverages elements from multi-armed bandits and game theory.
    Our approach has the advantage that it does not require a-priori knowledge of the MU's effort for each task and it incorporates the individual utility functions of the MUs and the MCSP.
    In contrast to existing works in this space~\cite{Xiao2017, Wang2018, Zhang2021, Gao2022, Xiao2021}, our approach considers the MUs and the MCSP to be individual rational decision makers.
    \item We propose an new ``free-sensing" mechanism to ensure that all MUs learn their expected effort for all task types thereby reducing future collisions. The idea behind the free-sensing strategy is that, occasionally, the MUs offer to perform tasks for free to ensure the tasks are assigned to them. 
    Performing a task for free is seen as an investment from the MU`s perspective, as the MU can improve its estimate of the required effort when performing said task.
    \item We show that the proposed decentralized \algoShort{} converges to a stable task assignment, where neither the MUs nor the MCSP have an incentive to change the task assignment. 
    Moreover, we prove that the stable regret, which is the expected loss incurred by not adopting the optimal assignment, is bounded by a sublinear function. 
    Additionally, we show that the computational complexity of the proposed decentralized online learning is only linearly dependent on the number of task types.
    \item We evaluate the performance of the proposed algorithm by comparing it with state-of-the-art baseline algorithms. The results verify that, under various settings, the proposed mechanism is effective in terms of worker satisfaction and MCSP's utility. Simulation results show that we achieve the optimum of the social welfare, which is the sum of the utility functions of MUs and the MCSP. Moreover, the proposed algorithm achieves an improvement of $16\,\%$ in terms of average task completion time compared to a state-of-the-art online learning algorithm. The performance is scalable and remains near-optimal even for large network sizes.
\end{itemize}

The rest of this paper is organized as follows. In Section~\ref{sec:system_model}, we introduce the MCS system model.
The proposed \algoShort{} is explained in Section \ref{sec:algorithm}.
In Section \ref{sec:analysisAlgorithm}, we analyze the offline optimal solution and prove that the proposed algorithm converges to a stable solution.
The numerical evaluation of \algoShort{} is presented in Section \ref{sec:numerical_evaluation} and finally, Section \ref{sec:conclusion} concludes the paper.

\vspace{-0.3cm}
\section{System model}
\label{sec:system_model}
\begin{table*}[t!]
\footnotesize
\centering
\caption{Table of notations}
\begin{tabular}{@{}cp{7cm} | cp{7cm}@{}}
\textsc{Symbol} & \textsc{Description} & \textsc{Symbol} & \textsc{Description}\\
\hline
    $\taskTypeZ$, $\numberOfTaskTypes$, $\setOfTaskTypes$ & Task type, Number of task types, Set of task types & 
    $\txPower$& Transmission power of $\MUk$ \\
    
    $\compPower$ & Power required for computation at $\MUk$ &
    $\taskSize$, $\taskComplexity$ & Size of task type $z$, Complexity of task type $z$\\
    
    $\taskWithIndex{n,t}$ & Task published at time $t$  & 
     $\payment$ & $\MUk$ earnings from task $\taskWithIndex{n,t}$ \\
    
    $\setOfTasks_\timeindex$ & Set of published tasks at time $t$ & 
    $\taskSize$ & Average result size for task type $\taskn$\\
    
    $\mappingFunction: \setOfTasks_\timeindex \rightarrow \setOfTaskTypes$ & Function mapping tasks to the type
    & $\totalEnergy$ & Energy used by $\MUk$ to complete $\taskWithIndex{n,t}$\\
    
    $\setOfTasksWithTypeZ$ & Set of all tasks with type $\taskTypeZ$ & 
    $\tau_{k,n,t}$ & $\MUk$ completion time for task $\taskWithIndex{n,t}$\\
    
    $\mathcal{I}$ & Complete information  & 
    $\deadlineOfTask$ & Average deadline of task type $\taskn$ \\
    
    $\rewardTaskCompletion$ & MCSP earnings from completion of task $\taskWithIndex{n,t}$ & 
    $\sensingTime$ & $\MUk$ sensing time for task $\taskWithIndex{n,t}$ \\
    
    $\mathcal{I}^{\mathrm{Task}}_n$,  $\mathcal{I}^{\mathrm{MU}}_k$ & MSCP-side, MU-side information & 
    $\communicationTime$ & $\MUk$ transmission time for task $\taskWithIndex{n,t}$ \\
    
	$K$ & Number of MUs  & 
	$\utilityMU$ & Utility of $\MUk$ for performing task $\taskWithIndex{n,t}$ \\
	
	$\mathcal{K}$ & Set of MUs  & 
	$\utilityTask$ & Utility of MSCP after task $\taskWithIndex{n,t}$ is performed \\
	
	$\mathbb{E}\{ X \}$ & Expected value of random variable $X$ & $\mathbb{P}(E)$ & Probability of event $E$ \\

\hline	\vspace{-0.8cm}
\end{tabular}
\label{tab:notation}
\end{table*}

We first describe our MCS system model.
A summary of the used notation is provided in Table \ref{tab:notation}.
We consider a set $\mathcal{K}$ of $K$ MUs who seek to perform tasks for the MCSP.
As shown in Figure~\ref{fig:system_model}, a single MCSP publishing $\numberOfTasks$ tasks is considered.
We consider a set $\setOfTaskTypes$ of $\numberOfTaskTypes$ different task types that represent several examples such as sensing temperature, taking a picture, or classifying an event.  
Each one of the $N$ tasks is classified according to their type $\taskTypeZ \in \setOfTaskTypes$. 
Time is divided into discrete time slots with index $\timeindex = 1,\dots, \timehorizon$. 
In each time slot $\timeindex$, the MCSP publishes a set of available tasks $\setOfTasks_t = \{\taskWithIndex{n,t}\}$, which can be seen in Fig.~\ref{fig:system_model_a}.
The mapping between task $\taskWithIndex{n,t}$ and its type $\taskTypeZ$ is given by a function $\mappingFunction: \setOfTasks_{\timeindex} \rightarrow \setOfTaskTypes$, i.e., $\mappingFunction(\taskWithIndex{n,t}) = \taskTypeZ$ means that $\taskWithIndex{n,t}$ is of type $\taskTypeZ$.
Furthermore, we collect all tasks of the same type $\taskTypeZ$ in the set $\setOfTasks_{\taskTypeIndex, \timeindex} \subseteq \setOfTasks_{\timeindex}$.
We assume that the MCSP may publish multiple tasks of the same type and each published task requires only one MU to complete.
The tasks are assumed to be time-sensitive by nature, i.e., the task's result must arrive in time at the MCSP~\cite{cheung2021timesensitive},~\cite{huang2022timedependent}.
Therefore, each task type $\taskTypeZ$ is characterized by the average size $\taskSize$ of its result, measured in bits, and an average deadline $\deadlineOfTask$.
The duration of the time slots is chosen according to the maximum completion time of a task.
We assume that the deadline $\deadlineOfTask$ is shorter than the duration of a time slot, i.e., tasks always have to be completed within one time slot.
Individual tasks $\taskWithIndex{n,t}$ of the same type $\taskTypeZ$ have different characteristics drawn from a type-specific, stationary probability distribution.
This probability distribution is unknown to the MCSP and the MUs.

The MCSP earns $\rewardTaskCompletion$ monetary units for the timely completion of a task $\taskWithIndex{n,t} \in \setOfTasksWithTypeZ$.
The earning $\rewardTaskCompletion$ is paid by the data requester.
To incentivize the MUs to participate, the MCSP pays the executing $\MUk$ when the task is finished before the deadline.
MUs are paid for the successful completion of the task according to the effort (time and energy) that $\MUk$ spent for the task completion~\cite{ahuja2019dynamic}.

\vspace{-0.5cm}
\subsection{Mobile Units}
In every time slot $\timeindex$, each $\MUk \in \setOfMUs$ can perform at most one task $\taskWithIndex{n,t}$.
Without loss of generality, we assume that every $\MUk$ is equipped with sensors that are capable of performing tasks from all $\numberOfTaskTypes$ task types.
To complete the assigned task, $\MUk$ has to spend effort in terms of time and energy.
The completion time $\totalTime$ of task $\taskWithIndex{n,t}$ contains three parts~\cite{huang2022timedependent}: the sensing time $\sensingTime$, the computation time $\computationTime$, and the communication time $\communicationTime$ for the transmission of the task's result.
The sensing time $\sensingTime$ is the time required by the MU to obtain valid sensing data. 
For example, in a traffic monitoring scenario, the platform requires $\MUk$ to record a specific-duration traffic video in a certain position of a road. 
The sensing time $\sensingTime$ of $\MUk$ for task $\taskWithIndex{n,t} \in \setOfTasksWithTypeZ$ is drawn from a stationary random distribution with the probability density function (PDF) $f^{\taskTypeZ}_{\sensingTime}(\sensingTime)$.
The expected value $\expectedSensingTime =\mathbb{E}(\sensingTime)$ of the sensing time depends on the task's type $\taskTypeZ$ and the $\MUk$ performing the task~\cite{huang2022timedependent}.

The computation time $\computationTime$ is the time required by $\MUk$ to preprocess the sensing data of a task of type $\taskTypeZ$.
Each MU is equipped with a central processing unit (CPU) with frequency $\localFrequency$.
The computation time is given by
\begin{align}
    \computationTime = \frac{\taskComplexity \taskSize}{\localFrequency},
\end{align}
whereas $\taskComplexity$ represents the preprocessing complexity of the task type $\taskTypeZ$.

The communication time $\communicationTime$ is the time required to transmit the preprocessed result of the task from $\MUk$ to the MCSP.
This time depends on the communication rate between $\MUk$ and the MCSP and it is drawn from a stationary random distribution with the PDF $f^{\taskTypeZ}_{\communicationTime}(\communicationTime)$.
The expected value $\expectedCommunicationTime =\mathbb{E}(\communicationTime)$ of the communication time depends on the size $\taskSize$ of the task result and the $\MUk$'s channel quality.
The total time $\MUk$ spends for task completion is $\totalTime = \sensingTime + \communicationTime + \computationTime$.
The time $\totalTime$ for task completion needs to be smaller than the deadline $\deadlineOfTask$.

Additionally, $\MUk$ must spend energy from its limited battery.
We assume that the energy $\totalEnergy$ used by $\MUk$ for the task completion is given by
    \begin{align}
        \totalEnergy = p_{k}^{\mathrm{comm}} \cdot \communicationTime + \compPower \cdot \computationTime,
    \end{align}
where $\txPower$ is the transmit power of $\MUk$ required to transmit the results of task $\taskWithIndex{n,t}$ and $\compPower$ is the power required for the computation.
We neglect the energy required for the sensors, as this energy consumption is small compared to the communication and computation energy~\cite{Capponi2017costEffective}.

In our model, all MUs have an MU-specific cost function $\costFunction(\totalTime,\totalEnergy)$ when performing a task. 
This cost function depends on the effort required to complete the task.
For example, some MUs may have a low battery level that results in a high cost to use energy $\totalEnergy$.
Other MUs might be concerned about the availability of their own communication, computation, or sensing resources, thus placing a high cost for the time $\totalTime$ during which the MU's resources are used.
We define the cost function as follows:
\begin{align}
\label{eq:costFunction}
        \costFunction(\totalTime,\totalEnergy) = \alpha_k \totalTime + \beta_k \totalEnergy.
    \end{align}
The cost function in $\eqref{eq:costFunction}$ captures the tradeoff between the completion time $\totalTime$ and the consumed energy $\totalEnergy$, with $\alpha_k$ being an MU-specific time cost parameter and $\beta_k$ an MU-specific energy cost parameter.

The MCSP pays $\payment$ monetary units to compensate $\MUk$ for the effort it spends to complete the task. This payment is defined as  

    \vspace{-0.6cm}
    \begin{align}
    \label{eq:paymentMU}
        \payment = \paymentFunction(\totalTime, \totalEnergy),
    \end{align}
where the payment function $\paymentFunction$ depends on the time and energy spent for the completion of the task.
The utility of $\MUk$ in time slot $\timeindex$ when performing task $\taskWithIndex{n,t}$ is
    \begin{align}
        \label{eq:utilityMU}
        \utilityMU = \payment \mathbbm{1}_{ \totalTime \leq \deadlineOfTask} - \costFunction(\totalTime,\totalEnergy),
    \end{align}
where $\mathbbm{1}_{\totalTime \leq \deadlineOfTask}$ is the indicator function for the case in which $\MUk$ completed the task before its deadline $\deadlineOfTask$.
The expected utility $\expectedUtilityMU$ for performing a task of type $\taskTypeZ$ is:
    \begin{align}
        \label{eq:expectedUtilityMU}
        & \expectedUtilityMU =  \mathbb{E}\{ \utilityMU | \taskWithIndex{n,t} \in \setOfTasksWithTypeZ\} 
         \\ &=   \mathbb{E}\{ \payment \} \cdot \mathbb{P}\{ \totalTime \leq \deadlineOfTask \}  
         - \mathbb{E}\{ \costFunction(\totalTime,\totalEnergy)\}. \nonumber 
    \end{align}

\vspace{-0.45cm}
\subsection{Mobile Crowdsensing Platform}
In each time slot $t$, the MCSP publishes a list of available tasks $\setOfTasks_t$ as shown in Fig.~\ref{fig:system_model_a}.
Each task from this list belongs to one of the $\numberOfTaskTypes$ task types.
The MCSP is paid by a data requester to provide results for each task $\taskWithIndex{n,t} \in \setOfTasks_t$.
The earning $\rewardTaskCompletion$ depends on the task type $\taskTypeZ$. 
Moreover, we assume $\rewardTaskCompletion$ to be deterministic and known beforehand to the MCSP, i.e., the MCSP and the data requester have made a contractual agreement.
The utility $\utilityTask$ of the MCSP when assigning $\MUk$ to task $\taskWithIndex{n,t} \in \setOfTasksWithTypeZ$ is defined as
    \begin{align}
        \label{eq:utilityMCSP}
        \utilityTask = (\rewardTaskCompletion - \payment) \mathbbm{1}_{\totalTime \leq \deadlineOfTask}.
    \end{align}
The expected utility $\expectedUtilityTask$ when assigning $\MUk$ to a task from task type $\taskTypeZ$ is given by
    \begin{align}
       \label{eq:expectedUtilityMCPS}
        \expectedUtilityTask &= \mathbb{E}\{ \utilityTask | \taskWithIndex{\taskIndex, \timeindex} \in \setOfTasksWithTypeZ\} \\ &= (\rewardTaskCompletion - \mathbb{E}\{ \payment \} )\cdot \mathbb{P}\{ \totalTime \leq \deadlineOfTask \}. \nonumber
    \end{align}

\subsection{Available information}
As the probability distributions $f^{\taskTypeZ}_{\communicationTime}(\communicationTime)$ and $f^{\taskTypeZ}_{\sensingTime}(\sensingTime)$ of the task characteristics are not known in advance, the MUs must estimate the average effort required for each task type.
We define $\mathcal{I}^{\mathrm{MU}}_k = \{ \expectedUtilityMU, \, \forall \taskTypeIndex \} $ as the \textit{MU-side} information about the stochastic characteristics of the task types, i.e., the average achievable utility $\expectedUtilityMU$ for each task type $\taskTypeZ$. 
$\mathcal{I}^{\mathrm{MU}}_k $ contains information about the expected energy consumption and the expected execution time for all task types $\taskTypeZ \in \setOfTaskTypes$. 
Note that $\mathcal{I}^{\mathrm{MU}}_k$ is not available at the MUs and has to be learned over time from experience.
Similarly, we define $\mathcal{I}^{\mathrm{Task}}_{\taskTypeIndex} = \{ \expectedUtilityTask, \, \forall k \} $ as the \textit{MCSP-side} information about the MUs. 
$\mathcal{I}^{\mathrm{Task}}_{\taskTypeIndex}$ contains information about the earnings and the required payment for all MUs.
As in the MU's case, $\mathcal{I}^{\mathrm{Task}}_\taskTypeIndex$ is not available at the MCSP in advance.
The combination of MU-side and MCSP-side information, $\mathcal{I} = \{ \mathcal{I}^{\mathrm{MU}}_\MUindex \cup \mathcal{I}^{\mathrm{Task}}_\taskTypeIndex, \, \forall \MUindex,\taskTypeIndex\} $, is called the \textit{complete} information and is unknown to the MUs and the MCSP.

Our goal is to optimize the assignment of tasks in a completely decentralized fashion without requiring prior knowledge of $\mathcal{I}$.
For this purpose, the MUs learn the characteristics of each task type and find their most preferred task in each time slot $t$.
In turn, the MCSP has to identify the best $\MUk$ to select for each task type.
We assume strict privacy constraints, meaning that the MUs do not share information about $\mathcal{I}^{\mathrm{MU}}_k$, neither with the MCSP nor with other MUs.
Additionally, the MCSP does not share $\mathcal{I}^{\mathrm{Task}}_\taskTypeIndex$ with the MUs.

We argue that a decentralized online learning strategy is an efficient solution to the task assignment problem.
Through online learning we can effectively address the key challenge of incomplete information.
Adopting a decentralized learning strategy ensures rigorous privacy for the MUs since they do not need to share their local information $\mathcal{I}^{\mathrm{MU}}_k$.
Moreover, a decentralized approach reduces the complexity of the problem.
This is because we can leverage the individual learning capabilities of each MU, thus eliminating the need to tackle the combinatorial problem at a centralized controller.
To analyze the task assignment problem from the perspective of the MUs and the MCSP, we first present the task assignment game between MUs and MCSP.
\vspace{-0.4cm}
\subsection{Problem Formulation: Task Assignment Game}
\label{sec:taskGame}
In contrast to either MU-centric MCS~\cite{cheung2021timesensitive}, or MCSP-centric MCS~\cite{guoju2020cmab}, we consider the perspective of both, the selfish MUs and the selfish MCSP.
Contrary to~\cite{cheung2021timesensitive} and~\cite{guoju2020cmab}, we do not formulate a global objective function for the performance of the task assignment.
Instead, we consider all MUs and the MCSP to be rational decision makers with their individual preferences and decision making capabilities.
Therefore, we use game theory, specifically matching theory~\cite{Gu2015matching}, to analyze the task assignment problem.
The main goal of matching theory is to obtain a stable matching, i.e., reaching a situation in which MUs and MCSP cannot simultaneously improve by changing the task assignment.
This corresponds to selfishly-deciding MUs and an MCSP that individually try to obtain their best task assignment.
A \emph{stable matching} outcome is apropos for the presented MCS problem because it allows the maximization of satisfaction for both the MUs and the MCSPs, with regard to their individual preferences.

The matching game is a model for a two-sided market in which the MUs provide their sensing resources and the MCSP requires sensing resources. 
These demands come in the form of indivisible sensing tasks, which the MUs execute in exchange of a payment~\cite{shapley1971assignment}.
The payment function $\paymentFunction$ and the MUs' cost function $\costFunction$ are given functions which depend on the task assignment~\cite{cen2022regret}.
The proposed, matching-based task assignment game $\mathcal{G}_t$ in time slot $t$ is formally described by a tuple $\mathcal{G}_t = (\setOfMUs, \setOfTasks_t, \MUpreference, \Taskpreference)$ containing the set $\setOfMUs$ of MUs, the set $\setOfTasks_t$ of available tasks, the MUs' preference ordering $\MUpreference$, and the MCSP's preference ordering $\Taskpreference$.
The MUs' preference ordering $\MUpreference$ ranks task types according to the expected utility associated with the task type $\taskTypeZ$, i.e., 

\vspace{-0.9cm}
    \begin{align}
        \taskTypeZ \MUpreference{} \taskTypeZ' \iff \expectedUtilityMUWithIndex{\MUindex,\taskTypeZ} \geq \expectedUtilityMUWithIndex{\MUindex,\taskTypeZ'}.
    \end{align}
In other words, $\MUk$ prefers task type $\taskTypeZ$ over $\taskTypeZ'$ if the MU's expected utility~\eqref{eq:expectedUtilityMU} of performing tasks of type $\taskTypeZ$ is higher than of tasks of type $\taskTypeZ'$.
The preference orderings $\MUpreference$ can only be correctly determined with the MU-side information $\mathcal{I}^{\mathrm{MU}}_k$.
The MCSP prefers MUs which yield the highest expected utility for each task type $\taskTypeZ$, i.e., 
    \begin{align}
        \MUwithIndex{k} \Taskpreference{}\MUwithIndex{l} \iff \expectedUtilityTaskWithIndex{k,\taskTypeIndex} \geq \expectedUtilityTaskWithIndex{l,\taskTypeIndex}.
        \label{eq:MCSP_preference}
    \end{align}
    
\noindent{}The expression in \eqref{eq:MCSP_preference} implies that when performing task type $\taskTypeZ$, the MSCP prefers $\MUk$ because it provides a higher utility compared to $\MUwithIndex{l}$.
This preference ranking can only be correctly determined with the MCSP-side information $\mathcal{I}^{\mathrm{MCSP}}$.

$\MUk$ signals its willingness to participate in any task of the type $\taskTypeZ$ by sending a sensing offer $\sensingOffer$ as shown in Fig.~\ref{fig:system_model_b}. 
Based on the received offers, the MSCP performs the assignment according to its preference ordering $\Taskpreference{}$ as depicted in Fig.\ref{fig:system_model_c}.
We denote the task assignment by the binary variable $\elementOfAssignmentMatrix$.
When $\elementOfAssignmentMatrix = 1$, $\MUk$ is assigned to task $\taskWithIndex{n,t}$. Otherwise, $\elementOfAssignmentMatrix = 0$.
The variables $\elementOfAssignmentMatrix$ associated to all MUs and tasks in time slot $t$ are collected in the matrix $\assignmentMatrix$.
\begin{definition}
A task assignment $\assignmentMatrix$ is \emph{unstable} if there are two MUs, $\MUwithIndex{k}$ and $\MUwithIndex{l}$, and two tasks, $\taskWithIndex{n,t}$ and $\taskWithIndex{m,t}$, such that:
(i) $\elementOfAssignmentMatrixWithIndex{k,n,t}=1$, i.e. $\MUwithIndex{k}$ is assigned to task $\taskWithIndex{n,t} \in \setOfTasksWithTypeZ$.\\
(ii) $\elementOfAssignmentMatrixWithIndex{l,m,t}=1$, i.e. $\MUwithIndex{l}$ is assigned to task $\taskWithIndex{m,t} \in \setOfTasksWithType{z'}$.\\
(iii) 
$\taskTypeZ' \MUstrictPreferenceWithIndex{k} \taskTypeZ$ and $\MUwithIndex{k} \TaskpreferenceWithIndex{z'} \MUwithIndex{l}$, i.e., $\MUwithIndex{k}$ strictly prefers the task with type $\taskTypeZ'$ over its current matched task of type $\taskTypeZ$, and the MCSP would profit more if the task of type $\taskTypeZ'$ is performed by $\MUwithIndex{k}$ instead of its current matched $\MUwithIndex{l}$.
\end{definition}
The pair $(\MUwithIndex{k}, \taskTypeZ')$ is called a blocking pair~\cite{nisan2007gameTheoryBook}, because both the $\MUk$ and the MCSP are unsatisfied with the current assignment.
The existence of the blocking pair $(\MUwithIndex{k}, \taskTypeZ')$ causes the matching $\assignmentMatrix$ to be unstable because $\MUwithIndex{k}$ could switch to $\taskWithIndex{m,t} \in \setOfTasksWithType{z'}$ and both, the $\MUwithIndex{k}$ and the task $\taskWithIndex{m,t}$ would obtain a more efficient matching and therefore a higher expected utility.

The assignment $\assignmentMatrix$ is said to be stable if no blocking pairs exist~\cite{nisan2007gameTheoryBook}. 
In such cases, no MU or task could change the assignment and improve their expected utilities.
In MCS, this means that each MU is assigned to its most preferred task while the MCSP selects its most preferred MU for each task.
Note that the stable matching may not be unique. There are, in fact,  potentially multiple solutions.
We denote the set of stable solutions as $\mathcal{X}^{\mathrm{stable}}$ and 
define $a^\mathrm{stable}_k$ as a stable task for MU $k$. The expected utility of this task is $\stableExpectedUtilityMU = \expectedUtilityMUWithIndex{\MUindex, a^\mathrm{stable}_k}$.
    
\vspace*{-0.2cm}
\section{Collision-Avoidance Multi-Armed Bandit with Strategic Free Sensing}
\label{sec:algorithm}
%
%
%
For most existing works on matching and assignment games, it is customary to use the so-called deferred acceptance algorithm (see Section~\ref{sec:offline-matching}) that guarantees convergence to a stable matching~\cite{roth2008deferred}. 
However, for our MCS problem, this approach would not be adequate because of several reasons.
First, the MUs do not know how much effort is required for each task type $\taskTypeZ$. 
Consequently, each MU has to learn its MU-side information $\mathcal{I}^{\mathrm{MU}}_k$ and its preferences by exploration.
Second, collisions with competing MUs occur while exploring different task types.
To avoid collisions and to ensure a good learning performance, a collision-avoidance mechanism is required.
As such, we propose a novel approach that combines online learning with matching theory including a collision-avoidance mechanism. 
This is more appropriate here because we can overcome the challenge of incomplete information and collisions due to the competition of the MUs. 

In each time slot $\timeindex$, $\MUk$ may send one sensing offer $\sensingOffer$ for a task type $\taskTypeZ$ together with its payment proposal $\MUpaymentProposal$.
The payment proposal $\MUpaymentProposal$ is calculated by the MUs based on their observed efforts for task type $\taskTypeZ$.
To lower the communication overhead between the MCSP and the MUs, we assume that the MUs can only send sensing offers for one task type at a time.
The MUs' challenge in sending a good sensing offer lies in the fact that the MUs do not know their expected utility and effort, i.e. time and energy, required to complete tasks of type $\taskTypeZ$ in advance.
When more MUs attempt to execute the same task type than tasks are available, i.e., sensing offers are colliding, the MCSP decides which MUs are assigned to the tasks according to the MCSP's utility~\eqref{eq:utilityMCSP} and the number $|\setOfTasksWithTypeZ|$ of tasks with type $\taskTypeZ$.
As shown in Fig.~\ref{fig:system_model_c}, the MCSP then sends a response $\sensingAccepted$ which contains whether the sensing offer was accepted, and which task was assigned to the MU.

Only the MU accepted by the MCSP and therefore, assigned to $\taskWithIndex{n,t}$, i.e., $\sensingAccepted = \taskWithIndex{n,t}$, can perform the task. Therefore, it is the only MU able to measure its utility $\utilityMU$ and effort in terms of time $\totalTime$ and energy $\totalEnergy$.
The MUs which were declined only learn that there are other MUs competing for task type $\taskTypeZ$ which were preferred by the MCSP. 
The competition between the MUs for the sensing tasks is especially challenging in the exploration phase, i.e., when the utility and effort for each task type are not well estimated. 
As a result, the payment proposals are either too low, which leads to a low utility, or too high, which increases the probability of a sensing offer being declined.

\begin{algorithm}[t]
\caption{CA-MAB-SFS (MUs' online learning)}\label{alg:proposed-algorithm}
\begin{algorithmic}[1]
\footnotesize
\REQUIRE $\epsilon_t, \lambda \in [0,1), \alpha \in [0,1)$
\STATE $\estimatedUtilityWithIndex{k,0}(\taskTypeZ) = 0 , \hat{J}_{k,0}(\taskTypeZ) = 0, \gamma_{k,\taskTypeZ} = 0 \; \forall k\in\mathcal{K}, \taskTypeZ\in\mathcal{Z}$
\FOR{$t = 1,\dots,T$}
    \STATE MCSP publishes sensing tasks $\setOfTasks_t$ and $P_{\taskTypeIndex,t-1} = \max \{ \MUpaymentProposal | x_{k,n,t-1} = 1, \taskWithIndex{n,t} \in \setOfTasksWithTypeZ\}$.
    \STATE Determine available task types $\setOfTaskTypes$ from the set $\setOfTasks_t$ of published tasks and the sets $\setOfTasksWithTypeZ$.
            \IF{$t = 1$}
                \STATE $\MUk$ sends sensing offer $\sensingOffer \gets \taskTypeZ, $ to a uniformly random chosen task type $\taskTypeZ \in \setOfTaskTypes$.
            \ELSE
                \STATE Draw i.i.d. random variable $D_{k,t}$ with $\mathbb{P}(D_{k,t} = 1) = \lambda$, $\mathbb{P}(D_{k,t} = 0) = 1-\lambda$.
                \IF{$D_{k,t} = 0$}
                    
                    \FOR{\textbf{each} $\taskTypeIndex \in 1,\dots, \numberOfTaskTypes $}
                    \STATE{\textbf{if} $\gamma_{k,\taskTypeIndex} > \freeSensingAdjustmentPara$ \textbf{then} $\MUpaymentProposal \xleftarrow[]{}0$}
                    \algorithmiccomment{free sensing offer}
                    \STATE{\textbf{else} $\MUpaymentProposal \xleftarrow[]{} \paymentFunction(\hat{J}_{k,t-1}(\taskTypeIndex))$} \algorithmiccomment{paid sensing offer}
                    \ENDFOR
                    \STATE 
                    Update plausible set, i.e., $S_k = \{ \taskTypeZ: P_{\taskTypeIndex,t-1} \geq \MUpaymentProposal,  \forall \taskTypeIndex=1,\dots,\numberOfTaskTypes\} $ 
                    \STATE Select $\taskTypeZ \in S_k$ using $\epsilon$ - greedy and send sensing offer $\sensingOffer \xleftarrow[]{} \taskTypeZ$.
                \ELSE
                    \STATE Send same sensing offer $\sensingOffer \gets \sensingOfferWithIndex{k,t-1}$ as in the previous timestep.
                \ENDIF 
            \ENDIF 
        \STATE{Wait for the MCSP's decision $\sensingAccepted$ from Algorithm~\ref{alg:proposed-algorithm-mcsp}.}
        \IF{$\MUk$ is accepted, i.e., $\sensingAccepted = \taskWithIndex{n,t}$}
            \STATE Assign the task to $\MUk$, i.e., $\elementOfAssignmentMatrixWithIndex{k,n,t} \gets 1$, where $\sensingAccepted = \taskWithIndex{n,t}$.
            \STATE Perform the task $\taskWithIndex{n,t}$ and observe $\utilityMU$, $\totalTime$ and $\totalEnergy$.
            \STATE Update estimates $\estimatedUtilityWithIndex{k,t}(\taskTypeIndex)$ and $\hat{J}_{k,t}(\taskTypeIndex)$.
            \STATE Reset rejection counter, i.e. $\gamma_{k,\taskTypeIndex} \xleftarrow[]{} 0$.
        \ELSE
            \STATE $\estimatedUtilityWithIndex{k,t}(\taskTypeIndex) \xleftarrow[]{}  \estimatedUtilityWithIndex{k,t-1}(\taskTypeIndex)$, $\hat{J}_{k,t}(\taskTypeIndex) \xleftarrow[]{}  \hat{J}_{k,t-1}(\taskTypeIndex)$
            \STATE{\textbf{if}\,\,$t<\freeSensingStopPara$}\,\,
            \textbf{then} {increase rejection counter of task type $\taskTypeZ$, i.e., $\gamma_{k,\taskTypeIndex} \xleftarrow[]{} \gamma_{k,\taskTypeIndex} + \frac{1}{t}$} 
        \ENDIF
    \ENDFOR  
\end{algorithmic}
\end{algorithm} 
In this section, our goal is to provide a fully decentralized online learning algorithm, which overcomes the challenges of the unknown information $\mathcal{I}$ and the competition between MUs.
In particular, we propose a novel decentralized online learning method termed CA-MAB-SFS.
The algorithm is fully decentralized and it consists of two strategies: The strategy of the MUs and the strategy of the MCSP.
The strategy of the MU is to select the best task type $\taskTypeZ$ for which to send a sensing offer and the payment proposal.
The strategy of the MCSP is to select the best sensing offers out of the received MUs' sensing offers for each task type.
Our algorithm only requires information exchange between the MUs and the MCSP. No information is exchanged between different MUs.

As mentioned before, a major challenge for the MUs is the exploration of task types, particularly at the beginning. Exploration is needed to estimate the effort associated with each task type. 
However, at the beginning, all MUs compete with each other because they all have only poor estimates of the required effort for each task type.
Intuitively, MUs may get rejected by the MCSP because they overestimated the effort associated with a task type.
This will cause high payment proposals for this task type in the future, leading to further rejections and, thus, to an inability to correctly learn the estimate of the effort. 
To overcome this, we propose the concept of \emph{strategic free sensing}.
MUs can decide to sense a task from a certain task type for free and in exchange learn about the task type characteristics.
This is done in the following way: The MU proposes to the MCSP to perform the task for free, i.e., the payment proposal $P_{k,n,t}$ is $0$.
Each $\MUk$ updates a rejection counter $\gamma_{k,\taskTypeZ}$ for each task type $\taskTypeZ$ if it has been rejected by the MCSP.
After a threshold value is reached, the MU sends a free sensing offer to get accepted with a high probability.

Algorithm~\ref{alg:proposed-algorithm} describes the online learning process of each MU.
In each time slot $t$, $\MUk$ receives a list of available sensing tasks $\setOfTasks_t$ together with information about the payment proposal 
\begin{align}
    \label{eq:paymentLastRound}
    P_{\taskTypeIndex,t-1} = \max \{ \MUpaymentProposal | x_{k,n,t-1} = 1, \taskWithIndex{n,t} \in \setOfTasksWithTypeZ\}
\end{align}
of the MU which was most expensive in the previous task assignment in $t-1$ for each task type (line 3).
In the first time slot $\timeindex = 1$,  $\MUk$ sends a sensing offer for a random task type (line 5-7), as no information about the utility and the effort for each task type is available.
For $t > 1$, $\MUk$ draws a random number $D_{k,t}$ which is equal to one with probability $\lambda$ and zero with probability $1-\lambda$ (line 8).
If $D_{k,t} = 1$, $\MUk$ sends a sensing offer to the same type as in the offer sent in the last time slot $t-1$ (lines 16-18).
The idea behind this mechanism is that not all MUs change their sensing offers simultaneously, which is required for the convergence of the online learning~\cite{liu2021bandit}.
The parameter $\lambda$ controls the trade-off between initial learning speed and convergence, which is discussed in Section~\ref{sec:numerical_evaluation}.
If $D_{k,t} = 0$, $\MUk$ determines the payment proposal for each task type $\taskTypeZ$ based on its effort estimate $\hat{J}_{k,t}(\taskTypeZ)$.
If $\MUk$'s rejection counter $\gamma_{k,\taskTypeZ}$ is larger than a predefined threshold $\freeSensingAdjustmentPara$ (line 12), $\MUk$ offers to sense the task for free.
Furthermore, $\MUk$ determines the plausible set $S_k$  containing all task types $\taskTypeZ$ where its payment $\MUpaymentProposal$ is lower than $P_{\taskTypeIndex,t-1}$ from~\eqref{eq:paymentLastRound}, i.e., all the task types which 
$\MUk$ can perform for a lower or equal payment than the most expensive MU who performed a task of the same type in the last assignment (line 14).
A task type from the plausible set $S_k$ is chosen according to the $\epsilon$-greedy strategy~\cite{auer2002finite}, i.e. with probability $\epsilon_t$ a random task type is chosen, and with probability $1-\epsilon_t$ the task with the highest expected utility is selected (line 15).
The sensing offer $\sensingOffer$ with the payment proposal $\payment$ is sent to the MCSP.
Afterwards, $\MUk$ waits for the response of the MCSP, described in Algorithm \ref{alg:proposed-algorithm-mcsp}.

After $\MUk$ receives the MCSP's response, its next action depends on whether it was accepted or not.
If $\MUk$ was accepted, the task $\taskWithIndex{n,t}$ is performed and the utility $\utilityMU$ and the effort regarding time $\totalTime$ and $\totalEnergy$ is observed and used to update the estimate $\estimatedUtilityWithIndex{k,t}(\taskTypeZ)$ of the utility and the estimate $\hat{J}_{k,t}(\taskTypeZ)$ of the task types's effort.
The update of $\estimatedUtilityWithIndex{k,t}(\taskTypeZ)$ is then given as

\vspace*{-0.5cm}
\begin{align}
    \estimatedUtilityWithIndex{k,t}(\taskTypeZ) = \estimatedUtilityWithIndex{k,t-1}(\taskTypeZ) + \frac{1}{N_k(z)} \cdot (\utilityMU - \estimatedUtilityWithIndex{k,t-1}(\taskTypeZ)),
\end{align}\vspace*{-0.4cm}

which is the iterative estimate of the mean value of $\utilityMU$, where $N_k(z)$ represents the number of times that MU $k$ has been assigned to task type $z$.
The estimate of the effort $\hat{J}_{k,t}(\taskTypeZ)$ for task type $\taskTypeZ$ is updated analogously.
If $\MUk$ was rejected by the MCSP, it receives no information about the utility of the task type and the required effort (line 26).
Only the rejection counter $\gamma_{k,\taskTypeZ}$ of task type $\taskTypeZ$ is increased by the value $t/\freeSensingSensitivityPara$ (line 27).
The analysis of the convergence of the proposed CA-MAB-SFS is presented in the following Section~\ref{sec:analysisAlgorithm}.

Algorithm~\ref{alg:proposed-algorithm-mcsp} describes the decision-making process of the MCSP for each task.
After the list of available tasks is published by the MCSP, it waits for the MUs' sensing offers.
Then, for each task type, the MCSP selects the MUs with the lowest payment proposal to complete all $|\setOfTasksWithTypeZ|$ tasks of type $\taskTypeZ$ (line 5).
If the lowest payment proposal is larger than $\rewardTaskCompletion$, the MCSP rejects all MUs.
For each $\MUk$, the MCSP sends a response $\sensingAccepted$ indicating whether the MU is accepted or rejected.
 \begin{algorithm}[t]
\caption{CA-MAB-SFS (MCSP's decision)}\label{alg:proposed-algorithm-mcsp}
\begin{algorithmic}[1]
\footnotesize
\REQUIRE $\setOfMUs, \setOfTasks, \rewardTaskCompletion$
\FOR{$t = 1,\dots,T$}
    \STATE Publish available sensing tasks $\setOfTasks_{t}$ and $P_{\taskTypeIndex,t-1} = \max \{ \MUpaymentProposal | x_{k,n,t-1} = 1, \taskWithIndex{n,t} \in \setOfTasksWithTypeZ\}$.
    \STATE Wait for all sensing offers $\sensingOffer$ and payment proposals $\MUpaymentProposal$.
    \FOR{$\taskTypeIndex = 1,\dots,\numberOfTaskTypes$}
        \STATE Select the $|\setOfTasksWithTypeZ|$ MUs with the lowest payment proposals.
        \STATE Send acceptance response to the selected MUs, i.e., $\bar{O}_{k,\timeindex} = \taskWithIndex{n,t} \; \forall \taskWithIndex{n,t} \in \setOfTasksWithTypeZ$
        \STATE Send rejection response to all other MUs, i.e., $\bar{O}_{l,\timeindex} = \varnothing$.
    \ENDFOR
\ENDFOR
\end{algorithmic}
\end{algorithm}

\vspace*{-0.4cm}
\section{Convergence and Regret Bound Analysis for the Proposed CA-MAB-SFS Algorithm}
\label{sec:analysisAlgorithm}
In this section, we show that the proposed algorithm is guaranteed to converge to a stable solution and its regret bound is fixed.
For the proof, we assume that in each round the number $|\setOfTasksWithTypeZ|$ of tasks of each type is fixed.
Furthermore, we assume that the mapping function $\mappingFunction: \setOfTasks_{t}\rightarrow \setOfTaskTypes$ is constant over time, i.e., the type of the task $\taskWithIndex{n,t}$ is the same in every round.
This applies to MCS scenarios in which each task has to be repeated regularly to update the measurements, e.g., traffic or temperature measurements in a smart city.

\vspace{-0.3cm}
\subsection{Solution with complete information}
\label{sec:offline-matching}

%
In this section, the solution of the matching-based, task assignment game is discussed when all players have complete information $\mathcal{I}$.
This assumption is unrealistic and it is only used to derive a baseline for our CA-MAB-SFS algorithm.
We define the oracle as a decision maker with complete information $\mathcal{I}$ who is able to calculate an stable solution in one time slot $t$. 
When every MU and the MCSP know $\mathcal{I}$, a stable solution of the task assignment game can be calculated using the deferred acceptance algorithm~\cite{roth2008deferred}.
\begin{algorithm}[t]
\caption{Offline Deferred Acceptance}\label{alg:deferred-acceptance}
\begin{algorithmic}[1]
\footnotesize
\REQUIRE $\mathcal{G}_t = (\setOfMUs, \setOfTasks_t, \MUpreference, \Taskpreference)$
\STATE Determine available task types $\setOfTaskTypes$ from the set $\setOfTasks_t$ of published tasks.
\STATE $\sensingOffer \xleftarrow[]{}  \varnothing, \; \mathcal{Z}^{\text{history}}_{k} \xleftarrow[]{}  \{\}, \; \forall k \in \mathcal{K}$
\WHILE {$\exists \sensingOffer = \varnothing \land \mathcal{Z}^\text{history}_{k} \neq \setOfTaskTypes$}
    \STATE Send sensing offer $\sensingOffer$ for task type $\taskTypeZ$, with $\taskTypeZ: \taskTypeZ \MUpreference \taskTypeZ', \; z \neq z', \; \forall z,z' \in \{\setOfTaskTypes \setminus \mathcal{Z}^\text{history}_{k}\}$
    \IF{ all $\taskWithIndex{n,t} \in \setOfTasksWithTypeZ$ are assigned }
        \IF{$\MUk \Taskpreference \MUwithIndex{l}$} 
            \STATE Assign task $\taskWithIndex{n,t}$ to $\MUk$ instead of $\MUwithIndex{l}$, i.e., $\elementOfAssignmentMatrixWithIndex{k,n,t}=1,\elementOfAssignmentMatrixWithIndex{l,n,t}=0$ 
        \ENDIF
    \ELSE
      \STATE $\mathbf{if}\,\, \MUk \Taskpreference \varnothing\,\, \mathbf{then}\,\,$ assign task $\taskWithIndex{n,t}$ to $\MUk$, i.e., $\elementOfAssignmentMatrixWithIndex{k,n,t}=1$ 
    \ENDIF
    \STATE $\mathcal{Z}^{\text{history}}_{k} \xleftarrow[]{}  \mathcal{Z}^{\text{history}}_{k} \cup \{\taskTypeZ\}$ \algorithmiccomment{Add task type $\taskTypeZ$ to the proposal history}
\ENDWHILE
\RETURN $\assignmentMatrix = \{\elementOfAssignmentMatrixWithIndex{k,n,t}\}_{\forall k,n} $
\end{algorithmic}
\end{algorithm}
The deferred acceptance algorithm to reach a stable task assignment is presented in Algorithm~\ref{alg:deferred-acceptance}.
The input is the task assignment game $\mathcal{G}_t$ in time slot $t$, where all players have access to the complete information $\mathcal{I}$.
Each MU is initialized without any assigned task and an empty sensing offer history $\mathcal{Z}^\text{history}_k$. 
After receiving the set $\setOfTasks_{t}$ from the MCSP, each MU determines the set $\setOfTaskTypes$ of available task types (line 1).
The sensing offer history $\mathcal{Z}^\text{history}_k$ contains all the task types $\taskTypeZ$ to which $\MUk$ has sent a sensing offer $\sensingOffer$ in the considered time slot $t$ (line 2). 
The algorithm is an iterative approach that runs as long as at least one MU remains unmatched and there are task types to which it has not yet sent a sensing offer (line 3). 
Each unmatched $\MUk$ sends a sensing offer considering its most preferred task type $\taskTypeZ$ which is not in the sensing offer history (line 4).
If all the tasks $\taskWithIndex{n,t}$ of type $\taskTypeZ$ are already assigned, and the sensing offer from $\MUk$ has a higher expected utility than any of the assigned $\MUwithIndex{l}$, the current assigned $\MUwithIndex{l}$ is exchanged with $\MUk$ (lines 5-9).
If there are still unassigned tasks of type $\taskTypeZ$, $\MUwithIndex{k}$ is assigned to one of these tasks as long as $\MUwithIndex{k}$ has a positive utility (line 11).
$\MUk$ adds the task type $\taskTypeZ$ to which it sent its sensing offer to its sensing offer history (line 13). 
When all MUs are either assigned to a task or have sent sensing offers to all task types, the output is a stable task assignment $\assignmentMatrix$. 
Note that Algorithm~\ref{alg:deferred-acceptance} is only used as a benchmark and cannot be implemented in real applications due to its strict requirement on $\mathcal{I}$, which as discussed before, cannot be fulfilled.

\vspace*{-0.35cm}
\subsection{Convergence and regret bound for CA-MAB-SFS}
In the decentralized task assignment setting, the \textit{stable regret} concept~\cite{liu2021bandit} is used to evaluate the performance of learning algorithms.
The stable regret describes the performance compared to the offline stable task assignment with complete information from Section~\ref{sec:offline-matching}.
We define the instantaneous stable regret in $t$ as

\vspace{-0.7cm}
\begin{align}
\label{eq:definitionInstantaneousStableRegret}
 r_k(t) = \stableExpectedUtilityMU - \sum_{\taskIndex=1}^{N} \elementOfAssignmentMatrix \expectedUtilityMU.
\end{align}
$r_k(t)$ is computed as the difference between the expected utility $\stableExpectedUtilityMU$ for the stable matching and the expected utilities of the task assignment $\assignmentMatrix$.
The stable regret of a sequence of task assignments $\{ \assignmentMatrix \}_{\timeindex = 1,\dots,\timehorizon}$ for $\mathrm{MU}_k$ is defined as 
\vspace*{-0.4cm}
\begin{align}
\label{eq:definitionStableRegret}
 R_k(T) = \sum_{\timeindex=1}^{\timehorizon} r_k(t).
\end{align}
$R_k(T)$ is computed as the sum of all instantaneous regrets over the whole time horizon $T$. 
\begin{theorem}
The stable regret is bounded by a sublinear function which is given by
\begin{align}
    R_k(T) \leq O \biggl( \Delta_k \frac{8\numberOfTaskTypes^5K^2 e^{^{\frac{\Delta^2}{\numberOfTaskTypes\umax{}}}}}{\rho^{\numberOfTaskTypes^4+1}(1-\frac{\Delta^2}{\numberOfTaskTypes\umax{}})} \log(T) T^{1-\frac{\Delta^2}{\numberOfTaskTypes\umax{}}} \biggr),
\end{align}
where $\rho = (1-\lambda)\lambda^{Z-1}$, $\Delta_k = \mathrm{max}_{\taskTypeIndex=1,\dots,\numberOfTaskTypes}\{ \stableExpectedUtilityMU - \expectedUtilityMU\}$ and $\Delta = \text{min}_{i,j \in , i \neq j} \{ \expectedUtilityMUWithIndex{k,i} - \expectedUtilityMUWithIndex{k,j} \} $.
\end{theorem}
\begin{proof}
See Appendix~\ref{app:proofTheorem1}.
\end{proof}
\vspace*{-0.3cm}
The stable regret $R_k(T)$ is bounded by a sublinear function, which means that the average instantaneous stable regret ${\overline{r}_k(t)= R_k(T)/T}$ goes to zero for $T\rightarrow\infty$.
The average instantaneous stable regret of the task assignment for each MU diminishes during the online learning procedure.

To prove the convergence of \algoShort, we analyze the probability $\mathbb{P}(\boldsymbol{X}_T \notin \mathcal{X}^{\mathrm{stable}})$ of not reaching a stable matching in time step $T$.
\begin{theorem}
The probability of not reaching a stable matching in time step $T$ is bounded by
\begin{align}
    \mathbb{P}(\boldsymbol{X}_T \notin \mathcal{X}^{\mathrm{stable}}) \leq O \biggl( \frac{8\numberOfTaskTypes^5K^2 e^{^{\frac{\Delta^2}{\numberOfTaskTypes\umax{}}}}}{\rho^{\numberOfTaskTypes^4+1}(1-\frac{\Delta^2}{\numberOfTaskTypes\umax{}})} \frac{\log(T)}{T^{\frac{\Delta^2}{\numberOfTaskTypes\umax{}}}} \biggr).
\end{align}
\label{proof:stability}
\end{theorem}
\begin{proof}
See Appendix~\ref{app:proofTheorem2}.
\end{proof}
This probability $\mathbb{P}(\boldsymbol{X}_T \notin \mathcal{X}^{\mathrm{stable}})$ goes to $0$ for $T\rightarrow\infty$ as $\lim_{T\rightarrow\infty} \frac{\log(T)}{T^{\frac{\Delta^2}{\numberOfTaskTypes\umax{}}}} = 0$.
This implies that the probability $\mathbb{P}(\boldsymbol{X}_T \in \mathcal{X}^{\mathrm{stable}})$ of achieving a stable matching approaches $1$, therefore CA-MAB-SFS converges.
When reaching a stable matching, all MUs and the MCSP would not profit from changing the assignment.

\vspace*{-0.3cm}
\subsection{Computational complexity analysis}
We now analyze the computational complexity of the proposed CA-MAB-SFS algorithm from the perspective of the MUs and the MCSP.
For the MUs, we analyze the complexity of one iteration of their learning algorithm (Algorithm \ref{alg:proposed-algorithm}). 
Note that the MU's decision only depends on the number $\numberOfTaskTypes$ of available task types. 
Therefore, we evaluate the algorithm's complexity with regard to $\numberOfTaskTypes$.
From Algorithm~\ref{alg:proposed-algorithm}, we can see that the complexity of lines 1-9 does not grow with the number $\numberOfTaskTypes$ of task types, therefore the computational complexity of each of this lines is constant and of the order $O(1)$. 
The complexity of line 10-15 is linearly dependent on the number of task types, as the loop iterates over each task type once, and therefore is of the order $O(\numberOfTaskTypes)$.
The lines 16-30 are not dependent on $\numberOfTaskTypes$ and are of constant complexity $O(1)$.
From this analysis, we can determine that the complexity of the proposed CA-MAB-SFS algorithm grows only linearly with the number $\numberOfTaskTypes$ of available task types, i.e., $O(\numberOfTaskTypes)$.

The MCSP has to choose among the set of proposing MUs $\setOfMUs$, and therefore the algorithm complexity is analyzed with regard to the number of MUs, $K$.
For the MCSP, the maximum computational complexity stems from the selection of the cheapest payment for each task (Algorithm~\ref{alg:proposed-algorithm-mcsp}, line 5).
For this, the MCSP has to evaluate the cost of each MU once, leading to a linear complexity with regard to the number $K$ of MUs.
Therefore, the computational complexity of the MCSP's algorithm is characterized by $O(K)$.

For both, the MUs and the MCSP, the communication overhead is low. 
The MCSP broadcasts the list of available tasks, receives the sensing offers and transmits the accept and defer messages.
Each MU only receives the list of available tasks, submits one sensing proposal, and receives an accept or defer message. 

\vspace*{-0.3cm}
\section{Simulation Results and Analysis}
\label{sec:numerical_evaluation}
\begin{table}
\footnotesize
\caption{Evaluation parameters}
\label{tab:evaluation_parameters}
	 \centering{
	 	\def\arraystretch{1.1}
	\begin{tabular}{|c|l|}
	\hline
	Parameter & Value  \\
	\hline
	Number of MUs & $K=100$ \\
	Size of the sensing task result~\cite{huang2022timedependent}& $\taskSize \in [ \SI{50}, \SI{100} ]~\SI{}{\mega\bit{}s}$\\ 
	Number of task types & $Z=10$ \\
	Tasks per task type & $|\setOfTasks_{\taskTypeIndex, \timeindex}|~\in  [5,10]$ \\
	Mean communication rate & $\expectedCommunicationTime \in [0.025,0.1] \SI{}{\frac{s}{\mega\bit}}$ \\
	CPU frequency~\cite{Mahn2021globalOrchestration}& $\localFrequency \in [1,2]~\SI{}{\giga\hertz}$ \\
    Mean sensing time~\cite{huang2022timedependent} & $\expectedSensingTime \in [60, 180]~\SI{}{\second}$ \\
	Transmission power~\cite{Mahn2021globalOrchestration}& $p_{k}^{\mathrm{comm}} = \SI{200}{\milli\watt}$ \\
	Power required for computation~\cite{Mahn2021globalOrchestration}& $p_{k}^{\mathrm{comp}} = \SI{1}{\watt}$ \\
	Computational complexity~\cite{huang2022timedependent} & $[200, 300]~\frac{\text{CPU Cycles}}{\text{bit}}$ \\
    \hline
	Earning of MCSP & $\rewardTaskCompletion = 1.4 + \SI{3} \cdot \taskSize$ \\ 
	MUs' cost for energy consumption & $\alpha_k =  \SI{0.01}{\frac{Monetary\,units}{\joule}}$ \\
	MUs' cost for time spent sensing  & $\beta_k =  \SI{0.004}{\frac{Monetary\,units}{\second}}$ \\
	Payment to the MUs & $\paymentFunction(\totalTime, \totalEnergy)$\\
	&$= 1.1 \cdot \costFunction(E_{k,n,t}, \totalTime)$ \\
	\hline
	Exploration rate& $\epsilon_t = \text{min}\{1 , 1 / t \}$\\
	Collision-avoidance parameter& $\lambda = 0.1$\\
	Free-sensing parameters & $\freeSensingStopPara = 30, \, \freeSensingAdjustmentPara  = 0.5$ \\ 
	\hline
	\end{tabular}}
\end{table}
In this section, we evaluate the performance of the proposed CA-MAB-SFS algorithm and compare it to baseline schemes.
\vspace*{-0.4cm}
\subsection{Evaluation metrics}
As the MUs and the MCSP have different goals, the assessment of the system's performance depends on the considered perspective.
We argue that different evaluation metrics need to be considered to assess the system's performance.
\subsubsection{Social Welfare}
Social welfare is often used in game theory to evaluate the performance of a solution from the whole network's perspective~\cite{Wang2019_Auction}. 
It is defined as the sum of all MUs' utilities and the MCSP's utility. 
\subsubsection{Average completion time}
We consider the time that is required to complete the tasks of the MCSP.
\subsubsection{Energy efficiency}
We consider the energy that is required to complete the tasks of the MCSP.
\subsubsection{Stability and number of blocking pairs}
Stability ensures that the MCSP and all MUs are satisfied, i.e., neither the MCSP nor the MUs have an incentive to deviate from the current task assignment.
Intuitively, stability is important to ensure that all MUs and the MCSP will use this strategy, as their individual goals are achieved~\cite{cen2022regret}.
The number of blocking pairs indicates how many MU-task pairs would profit from changing the task assignment.
We measure the number of MUs that are part of a blocking pair, which represents how many MUs could improve their utility by adopting another task assignment.

\begin{figure*}[t]
\minipage{0.32\textwidth}
  \includegraphics[width=\linewidth]{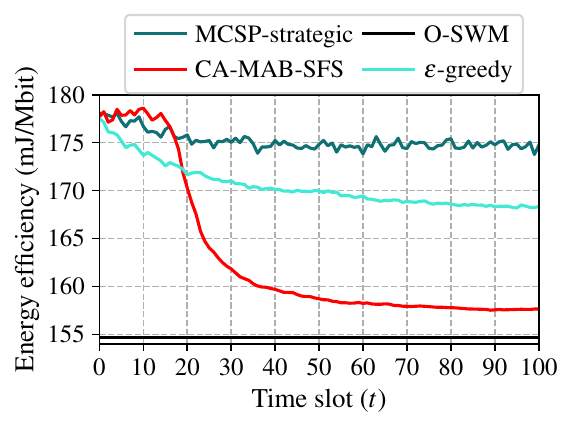}
  \vspace*{-0.8cm}
  \caption{Energy efficiency as a function of the time step $t$.}\label{fig:energy_efficiency}
\endminipage\hfill
\minipage{0.32\textwidth}
  \includegraphics[width=\linewidth]{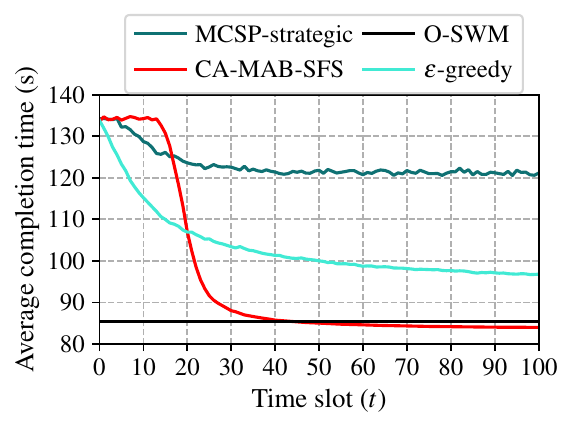}
  \vspace*{-0.8cm}
  \caption{Average task completion time as a function of the time step $t$.}\label{fig:completion_time}
\endminipage\hfill
\minipage{0.32\textwidth}
  \includegraphics[width=\linewidth]{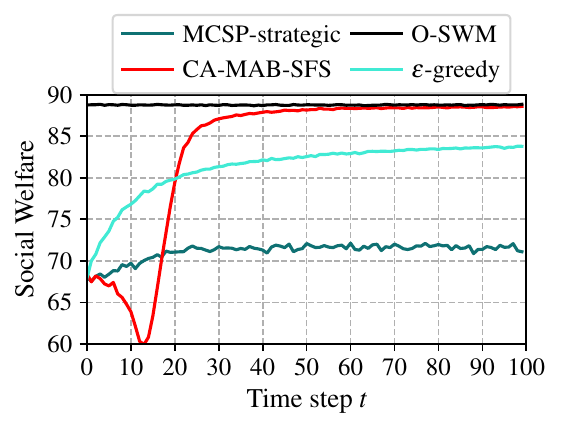}
  \vspace*{-0.8cm}
  \caption{Social welfare as a function of the time step $t$.}\label{fig:social_welfare}
\endminipage\hfill
\vspace{-0.4cm}
\end{figure*}
\begin{figure*}[t]
\minipage{0.32\textwidth}
  \includegraphics[width=\linewidth]{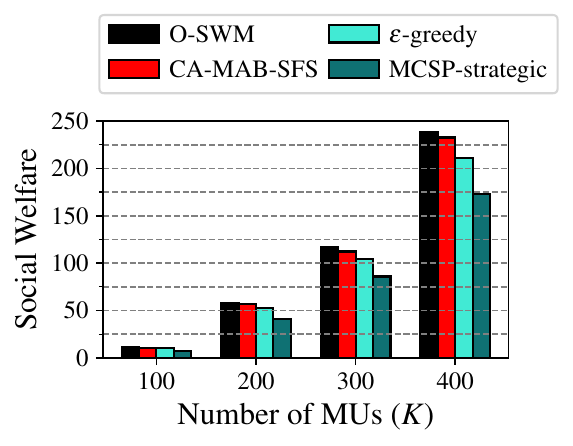}
  \vspace*{-0.8cm}
  \caption{Social welfare for an increasing network size for $K=N$, $Z=10$.}\label{fig:number_of_MUs}
\endminipage\hfill
\minipage{0.32\textwidth}
  \includegraphics[width=\linewidth]{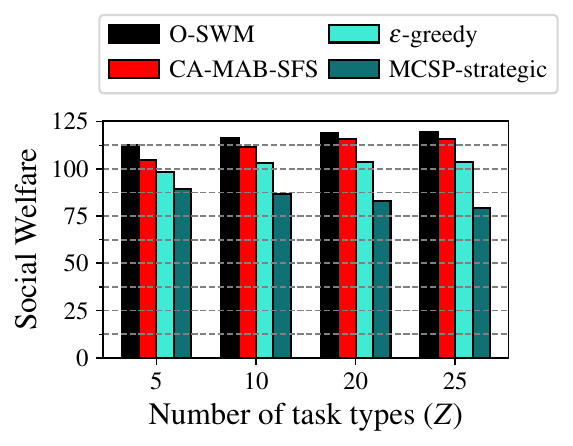}
  \vspace*{-0.8cm}
  \caption{Social welfare as a function of the number $\numberOfTaskTypes$ of task types, $K=N=100$.}\label{fig:number_of_task_types}
\endminipage\hfill
\minipage{0.32\textwidth}
  \includegraphics[width=\linewidth]{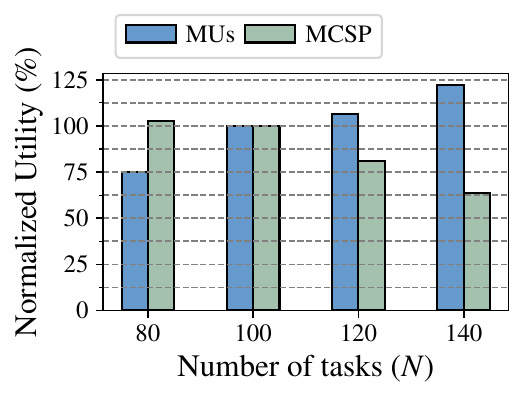}
  \vspace*{-0.8cm}
  \caption{Normalized utility of the MUs and the MCSP using CA-MAB-SFS as a function of $N$, $K=100$, $Z=20$.}\label{fig:competition}
\endminipage\hfil\vspace{-0.3cm}
\end{figure*}

\vspace{-0.2cm}
\subsection{Baseline Algorithms}
We use the following algorithms to benchmark our proposed CA-MAB-SFS. 
Assuming complete information $\mathcal{I}$ for each MU and the MCSP, we consider the following offline approaches:
\begin{itemize}
    \item \textit{Offline Deferred Acceptance Algorithm} (O-DAA), which is described in Section~\ref{sec:offline-matching} and Algorithm~\ref{alg:deferred-acceptance}.
    The complete information is available, therefore the payment of the MUs is calculated based on the actual effort required to perform the task, as specified in~\eqref{eq:paymentMU}.
    \item \textit{Offline Social Welfare Maximization} (O-SWM): Similar to the offline task assignment game in Section~\ref{sec:taskGame}, an optimization problem of the social welfare is formulated with complete information $\mathcal{I}$. The optimal solution is calculated using a solver from the OR-Tools~\cite{ortools}.
\end{itemize}
Additionally, we consider the following baseline algorithms which do not require complete information:
\begin{itemize}
  \item \textit{Decaying $\epsilon$-greedy online-learning}: Each MU uses the decaying $\epsilon$-greedy online-learning algorithm~\cite{auer2002finite} to learn the effort and utility for each task. 
  The exploration of tasks with high effort is performed according to a probability $\epsilon$ which is decreasing over time.
  In case the sensing offer of the MU is rejected, the MU's utility is assumed to be zero.
  \item \textit{Only MCSP-strategic}: Each MU randomly selects a task type $\taskTypeZ$ and sends a sensing offer to this task type. 
  The payment proposal is calculated using the average of the past efforts.
  The MCSP selects the MU with the lowest payment proposal.
\end{itemize}
\vspace{-0.3cm}
\subsection{Evaluation Setup}
For the simulations, the parameters listed in Table \ref{tab:evaluation_parameters} are considered, unless otherwise specified.
The number $K$ of MUs is chosen to be $K=100$.
The number $N$ of tasks is chosen from the interval $[50,100]$, whereas $Z=10$ different task types are available. 
%
%
The sensing time varies every time slot for each MU and is drawn from a normal distribution with mean $\expectedSensingTime$ and standard deviation $10\,\SI{}{\second}$.
%
%
The mean communication rate is randomly drawn from the interval $[10,40]\,\SI{}{\mega\bit\per\second}$, which corresponds to the mean communication time $\expectedCommunicationTime = [0.025,0.1]\,\SI{}{\second\per\mega\bit}\cdot\taskSize$.
The communication time varies in every time slot for each MU, and it is drawn from a normal distribution with mean $\expectedCommunicationTime$ and standard deviation $0.01\,\SI{}{\second\per\mega\bit}$.
The mean CPU frequency available at each MU is $\localFrequency \in [1,2]~\SI{}{\giga\hertz}$. Each time slot, it is drawn from a Gaussian distribution with the mean $\localFrequency$ and standard deviation $100~\SI{}{\mega\hertz}$.
For each figure, $100$ Monte-Carlo iterations were performed and the results are averaged.

\vspace{-0.3cm}
\subsection{Results and Discussion}

We assess the energy efficiency of the proposed CA-MAB-SFS algorithm and the baseline algorithms in Fig.~\ref{fig:energy_efficiency}.
The energy consumption is normalized to the size of the task result $\taskSize$, i.e., the energy efficiency is given by the energy consumed for each bit of the task result. 
The energy efficiency of the proposed CA-MAB-SFS is slightly lower than the baseline algorithms for $t < 20$.
This is due to the strategic free sensing mechanism in the CA-MAB-SFS algorithm, where MUs explore task types for free.
The MCSP prefers the MUs which perform the task for free over the most energy-efficient MUs, and therefore does not select the most efficient MU in this case. 
The exploration of task types is challenging due the competition between MUs, which initially causes poorer performance of the CA-MAB-SFS in the learning phase for $t < 20$.
When the exploration rate and the strategic free sensing reduces for $t > 20$, the CA-MAB-SFS shows a fast improvement in terms of energy efficiency.
Fig.~\ref{fig:energy_efficiency} demonstrates that for $t > 50$, the proposed CA-MAB-SFS algorithm achieves a $7.5\,\%$ increase in energy efficiency compared to the $\epsilon$-greedy algorithm and an $11.5\,\%$ increase compared to the MCSP-strategic algorithm.
Furthermore, the performance of the CA-MAB-SFS algorithm is within $1.2\,\%$ of the O-SWM algorithm which requires complete information. 

The average time required to complete the tasks is shown in Fig.~\ref{fig:completion_time}. 
For $t > 50$, the proposed CA-MAB-SFS algorithm outperforms the $\epsilon$-greedy by $16\,\%$ and the MCSP-strategic by $41\,\%$.
It achieves a slightly lower average task completion time than the O-SWM algorithm. 
This is due to the fact that the cost factor $\alpha_k$ of the MUs for the time is higher than the cost factor $\beta_k$ for the energy, therefore the MUs prefer to execute tasks which require a lower completion time.
The O-SWM algorithm maximizes the social welfare and therefore assigns tasks to MUs without considering their individual preferences, which will not yield the time-optimal result.
Initially, the CA-MAB-SFS algorithm is slightly worse than the baseline algorithms due to the strategic free sensing procedure, but then outperforms the baseline algorithms significantly. 

%
\begin{figure*}[t]
\minipage{0.32\textwidth}
  \includegraphics[width=\linewidth]{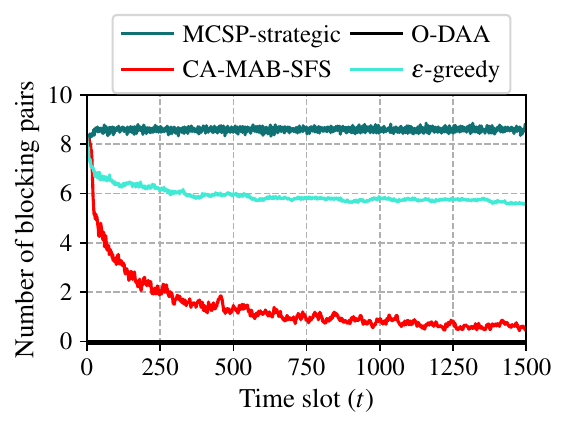}
  \vspace*{-0.8cm}
  \caption{Number of blocking pairs as a function of the time step $t$, $K=10,N=10,Z=10$.}\label{fig:blocking_pairs}
\endminipage\hfill
\minipage{0.32\textwidth}
  \includegraphics[width=\linewidth]{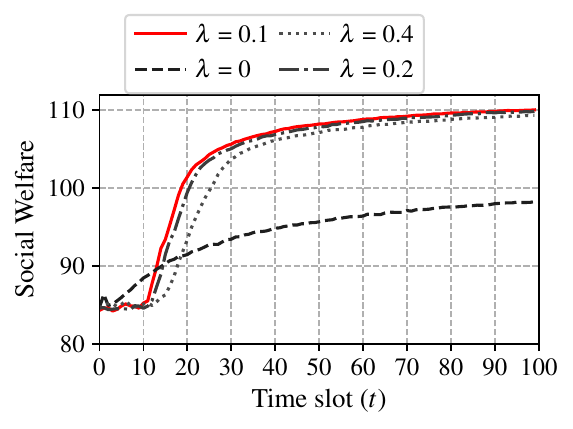}
  \vspace*{-0.8cm}
  \caption{Social welfare as a function of the time step $t$ for varying collision-avoidance parameters $\lambda$.}\label{fig:collision_avoidance_parameter}
\endminipage\hfill
\minipage{0.32\textwidth}
  \includegraphics[width=\linewidth]{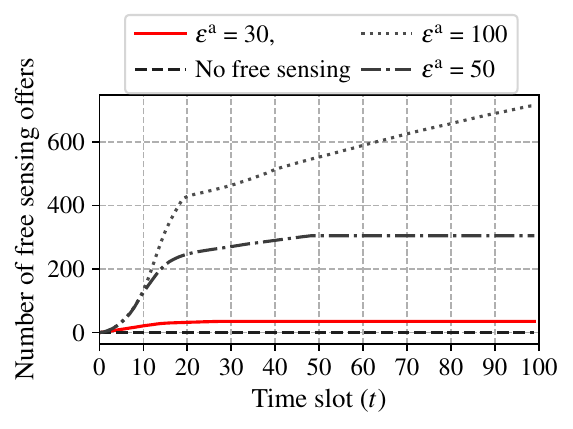}
  \vspace*{-0.8cm}
  \caption{Number of free sensing offers as a function of time step $t$ for varying SFS parameters.}\label{fig:free_sensing_offers}
\endminipage\hfill\vspace{-0.3cm}
\end{figure*}

Figure~\ref{fig:social_welfare} depicts the achieved social welfare of the different algorithms.
The achievable maximum of the social welfare is given by the task assignment of the O-SWM algorithm.
The proposed CA-MAB-SFS shows a good convergence to the social welfare maximum, whereas the $\epsilon$-greedy and the MCSP-strategic algorithm are not able to converge to the optimum.
The $\epsilon$-greedy online learning achieves a $7.2\,\%$ lower social welfare than the optimum and the MCSP-strategic a $22\,\%$ lower social welfare.
As in the Fig.~\ref{fig:completion_time}, the decrease in social welfare of the proposed CA-MAB-SFS for $t<20$ is due to the strategic free sensing mechanism, where some MUs execute tasks for free to learn more about the different task types.

The impact of the number $\numberOfMUs$ of MUs and the number $\numberOfTasks$ of tasks on the social welfare is shown in Fig.~\ref{fig:number_of_MUs} for $t = 1000$.
It can be seen that even for large MCS network sizes $K=N=400$, the proposed CA-MAB-SFS is within $2\,\%$ of the optimal social welfare given by the O-SWM algorithm, whereas the $\epsilon$-greedy achieves $14\,\%$ less social welfare.
For larger networks, CA-MAB-SFS achieves near-optimal social welfare, while $\epsilon$-greedy is $18\,\%$ below optimum.

The impact of the heterogeneity of tasks is shown in Fig.~\ref{fig:number_of_task_types}.
The number $Z$ of task types is varied while keeping the number $K$ of MUs and the number $N$ of tasks constant.
We observe for all values of $Z$ that the proposed algorithm achieves a near optimal performance within $6\,\%$ of the social welfare optimum. 

Next, we analyze the effect of the competition between the MUs by varying the ratio $K/N$ between the number of tasks and the number of MUs.
The utility of the MUs and the MCSP using CA-MAB-SFS is shown in Fig.~\ref{fig:competition} for a varying ratio between MUs and tasks.
For an increased competition between the MUs, i.e., less MUs than tasks ($K/N < 1$), we can see that the utility of the MUs decrease whereas the MCSP's utility increases.
This is due to the fact that MUs with a lower payment proposal are selected by the MCSP and therefore the average payment between for each task decreases.
When fewer MUs compete ($K/N > 1$), the utility of the MUs increases as they more frequently select tasks with higher payments.

To assess the stability of the solution, we depict the number of blocking pairs in Figure~\ref{fig:blocking_pairs}.
Note that fewer blocking pairs result in more MUs satisfied with the task assignment.
The proposed algorithm converges to zero blocking pairs, which demonstrates that CA-MAB-SFS converges to the stable solution, as shown in Theorem~\ref{proof:stability}.
The regular $\epsilon$-greedy algorithm, which does not consider the competition between the MUs, leaves $60\,\%$ of MUs that could improve by changing the task assignment. 
As the MCSP-strategic algorithm does not consider the utility of the MUs, more than $80\,\%$ of the MUs are not satisfied with the task assignment.

To understand the impact of the collision-avoidance parameter $\lambda$ on the performance, we varied $\lambda$ in Fig.~\ref{fig:collision_avoidance_parameter}.
For $\lambda = 0$, we observe a faster initial learning for $t<12$. This is due to the fact that the MUs' suboptimal decisions are not repeated.
However, this configuration does not converge to the maximum social welfare.
For $\lambda = 0.4$, we observe a significantly lower learning speed, as $40\,\%$ of the MUs in average repeat the same decision as in the last time slot, which is ineffective.
The collision-avoidance parameter therefore controls the trade-off between initial learning speed and convergence. 
Lower values of $\lambda$ exhibit a higher initial learning speed, but may converge much slower. 
Higher values of $\lambda$ have a lower initial learning speed, but converge faster.
For our simulations, we chose $\lambda = 0.1$ as it empirically yields the best results.

Fig.~\ref{fig:free_sensing_offers} shows the cumulative number of free sensing offers, i.e., how many sensing offers without payment proposal have been sent.
To analyze the impact of the free sensing parameter $\epsilon^{a}$, we analyze the cumulative number of free sensing proposals.
In our analysis of CA-MAB-SFS, we clearly see two phases: The phase with free sensing offers $t \leq \freeSensingStopPara$, and the phase without free sensing offers $t > \freeSensingStopPara$.
In the phase with free sensing offers, MUs submit a free sensing offer after their respective rejection threshold $\freeSensingAdjustmentPara$ is exceeded. 
This is done to ensure that the MU will be accepted by the MCSP and the effort estimate will improve.
From Fig.~\ref{fig:free_sensing_offers}, we can see that increasing $\epsilon^{a}$ leads to a higher number of cumulative free sensing offers. 
The number of free sensing offers does not increase for $t > \epsilon^{a}$, so it does not increase indefinitely.
\vspace{-0.35cm}
\section{Conclusion}
\label{sec:conclusion}
In this paper, we have studied the assignment of tasks in MCS. 
We have analyzed the conflicting interests of the MCSP and the MUs, the statistical nature of the tasks and MU's characteristics, as well as the competition between MUs.
To consider the conflicting goals of the MCSP and MUs, we have formulated a matching-based task assignment game.
We have proposed a novel decentralized online learning algorithm for the task assignment game, termed CA-MAB-SFS, which incorporates an innovative free sensing strategy.
We have then proven its convergence to a stable task assignment, i.e., an assignment where neither the MUs nor the MCSP can improve.
The stable regret, i.e., the loss of the online learning compared to having complete information, is bounded by a sublinear function and decreases to zero.
Furthermore, we showed that the computational complexity for each MU and the MCSP is low.
Simulation results show that, compared to the popular $\epsilon$-greedy online learning approach, our proposed CA-MAB-SFS algorithm does not only reduce the average completion time of tasks by $16\,\%$, but also enhances the energy efficiency of the MCS system by up to $7.5\,\%$.
We have also showed that the number of blocking pairs, i.e., the number of MUs that would improve by deviating from the task assignment, converges to zero.
Furthermore, we have proven that our proposed CA-MAB-SFS converges to the maximum of the social welfare, whereas state-of-the-art online learning approaches are not able to reach it. 
%
%
\vspace{-0.2cm}
\begin{spacing}{0.95}
\bibliographystyle{./bibliography/IEEEtran}
\bibliography{./bibliography/IEEEabrv,./bibliography/references}
\end{spacing}
\begin{appendices}
\vspace{-0.3cm}
\section{Proof of Theorem 1}
\vspace{-0.2cm}
\label{app:proofTheorem1}
The core idea is to bound the probability $\mathbb{P}(\mathbf{x}_t \notin X^{\mathrm{stable}})$ of the event that no stable matching is achieved until time $T$.
If a stable matching is reached, the stable regret of all MUs will be zero, otherwise, we bound the regret by the maximum regret over all MUs which is given by 
\begin{equation}
\Delta_k = \mathrm{max}_{\taskTypeIndex=1,\dots,\numberOfTaskTypes}\{ \stableExpectedUtilityMU - \expectedUtilityMU\}.
\end{equation}
Therefore, we formulate the following stable regret bound:
\begin{align}
    R_k(T) \leq \Delta_k  \sum_{t=0}^T \mathbb{P}(\mathbf{x}_t\notin{}X^{\mathrm{stable}})
\end{align}
As $\Delta_k$ can be directly calculated from the MUs expected utilities $\expectedUtilityMU$, we only need to bound the probability of an unstable matching $\mathbb{P}(\mathbf{x}_t \notin X^{\mathrm{stable}})$. 
The following events $E_{1,\timeindex}$ and $E_{2,\timeindex}$ prevent a stable matching:
\begin{itemize}
    \item $E_{1,\timeindex}$: At least one user is exploring according to $\epsilon$-greedy and not selecting its stable task $\stableTaskForMUk$ in $\timeindex$.
    \item $E_{2,\timeindex}$: Either one MU has statistical ranking mistakes, i.e. its estimates of the utility result in a sensing offer for a suboptimal task type, or there were no statistical ranking mistakes but the matching at time $t-1$ was unstable. 
\end{itemize}
Considering $E_{1,\timeindex}$ and $E_{2,\timeindex}$, we use the following bound
\begin{align}
    \label{eq:probabilityNotStable}
    \mathbb{P}(\mathbf{x}_t \notin X^{\mathrm{stable}}) & \leq \mathbb{P}(E_{1,\timeindex}) + \mathbb{P}(\overline{E}_{1,\timeindex})\mathbb{P}(E_{2,\timeindex}),
\end{align}
with $\overline{E}_{1,\timeindex}$ as the complementary event of $E_{1,\timeindex}$.
The stable regret bound is given by
\begin{align}
    R_k(T) 
    & \leq \Delta_k \biggl( \sum_{\timeindex=1}^T \mathbb{P}(E_{1,\timeindex}) +  \sum_{\timeindex=1}^T \mathbb{P}(E_{2,\timeindex}) \biggr),
\end{align}
using $\mathbb{P}(\overline{E}_{1,\timeindex}) \leq 1$.
In the following, we derive the probability of the events $E_{1,\timeindex}$ and $E_{2,\timeindex}$ separately. \\
\textbf{Derivation of $\mathbb{P}(E_{1,\timeindex})$}:
The probability of $E_{1,\timeindex}$ is given by
\vspace{-0.1cm}
\begin{align}
    \mathbb{P}(E_{1,\timeindex}) &\leq 1 - \biggl((1-\epsilon_t) + \epsilon_t\frac{1}{\numberOfTaskTypes}\biggr)^K \nonumber \\
                    &= 1 - \biggl(1-\epsilon_t\frac{\numberOfTaskTypes-1}{\numberOfTaskTypes}\biggr)^K 
\end{align}
which is the complementary event of all $K$ MUs exploiting in time $\timeindex$ or randomly selecting the stable task out of the $N$ tasks.
The summation over all time slots $t$ yields
\begin{align}
    \sum_{t=1}^{T}\mathbb{P}(E_{1,\timeindex}) & \leq \sum_{t=1}^{T} 1 - \biggl(1-\epsilon_t\frac{\numberOfTaskTypes-1}{\numberOfTaskTypes}\biggr)^K \nonumber\\
                                               &= T - \sum_{t=1}^{T} \biggl(1-\epsilon_t\frac{\numberOfTaskTypes-1}{\numberOfTaskTypes}\biggr)^K .
\end{align}
We bound the inner term of the sum by Bernoulli's inequality
\begin{align}
\biggl(1-\epsilon_t\frac{\numberOfTaskTypes-1}{\numberOfTaskTypes}\biggr)^K \geq 1 + K\cdot \epsilon_t\frac{\numberOfTaskTypes-1}{\numberOfTaskTypes},
\end{align}
that holds for $\epsilon_t\frac{\numberOfTaskTypes-1}{\numberOfTaskTypes} \leq 1$, which can be easily checked.
Therefore,
\begin{align}
\sum_{t=1}^{T}\mathbb{P}(E_{1,\timeindex})  &\leq \sum_{t=1}^{T}K\cdot \epsilon_t\frac{\numberOfTaskTypes-1}{\numberOfTaskTypes}
\end{align}
For a sufficiently fast decaying $\epsilon_t$, e.g. $\epsilon_t = \text{min}\{1,1/t\}$, we can show that:
\begin{align}
 \sum_{t=1}^{T}\mathbb{P}(E_{1,\timeindex}) &\leq K\frac{\numberOfTaskTypes-1}{\numberOfTaskTypes} \sum_{t=1}^{T}\text{min}\biggl(1,\frac{1}{t}\biggr) \nonumber \\
                                            &\leq K\frac{\numberOfTaskTypes-1}{\numberOfTaskTypes} (\log(T) + 1). \label{eq:p_e1t}
\end{align}
%
\textbf{Derivation of $\mathbb{P}(E_{2,\timeindex})$}:
Note that $\mathbb{P}(E_{2,\timeindex})$ is the probability of MUs having statistical ranking mistakes or not achieving a stable matching.
For this proof, we assume that, as in real applications, the utility of the MUs is limited to finite values in an interval $[U_{\mathrm{min}},U_{\mathrm{max}}]$.
Therefore, we can define $\umax = U_{\mathrm{max}} - U_{\mathrm{min}}$.
The proof is an adapted version of the proof in~\cite{liu2021bandit}, using arguments for $\epsilon$-greedy MABs from~\cite{auer2002finite}.
The authors of~\cite{liu2021bandit} show that
\begin{align}
\label{eq:p_e2t}
\sum_{t=1}^T\mathbb{P}(E_{2,\timeindex}) \leq 4\frac{\numberOfTaskTypes^5K^2}{\rho^{\numberOfTaskTypes^4+1}}\log(T)( x_0 + 12),
\end{align}
where $x_0 = \sum_{t=1}^T P\{\hat{Q}_{k,t}(i) > \hat{Q}_{k,t}(j) \cap x_{k,n,t} = 1\} \leq \sum_{t=1}^T P\{\hat{Q}_{k,t}(i) > \hat{Q}_{k,t}(j)\}$ denotes the expected number of statistical ranking mistakes up to $T$ and $\rho = (1-\lambda)\lambda^{Z-1}$.

The expected number $k_t$ of sensing offers to a suboptimal task type in time slot $t$ during the exploration phase is defined as
\begin{align}
\label{eq:kt}
k_t = \frac{1}{\numberOfTaskTypes}\sum_{t'=1}^{t} \epsilon_{t'} \leq \frac{1}{\numberOfTaskTypes}(\log (t) + 1).
\end{align}
Using Hoeffding's inequality and the definition in \eqref{eq:kt}, we can show that 
\begin{align}
\mathbb{P} \{ \hat{Q}_{k,t}(i) > \hat{Q}_{k,t}(j) \} &\leq 2 e^{-\frac{k_t \Delta^2}{\umax}} \nonumber\\
                                                     &= 2 e^{^{\frac{\Delta^2}{\numberOfTaskTypes\umax{}}}} \frac{1}{t^{\frac{\Delta^2}{\numberOfTaskTypes\umax{}}}},
\end{align}
with $\Delta = \text{min}_{i,j \in , i \neq j} \{ \expectedUtilityMUWithIndex{k,i} - \expectedUtilityMUWithIndex{k,j} \} $.
The expected number $x_0$ of statistical ranking mistakes up to $T$ can be bounded by 
\begin{align}
x_0 &\leq 2 e^{^{\frac{\Delta^2}{\numberOfTaskTypes\umax{}}}} \sum_{t=1}^{T}  \frac{1}{t^{\frac{\Delta^2}{\numberOfTaskTypes\umax{}}}} 
\leq 2 e^{^{\frac{\Delta^2}{\numberOfTaskTypes\umax{}}}} \int_{t=0}^{T}  \frac{1}{t^{\frac{\Delta^2}{\numberOfTaskTypes\umax{}}}}\, \mathrm{d}t \nonumber \\
& = \frac{2e^{^{\frac{\Delta^2}{\numberOfTaskTypes\umax{}}}}}{1-\frac{\Delta^2}{\numberOfTaskTypes\umax{}}}  T^{1-\frac{\Delta^2}{\numberOfTaskTypes\umax{}}} \label{eq:x0}.
\end{align}
Using \eqref{eq:p_e1t},~\eqref{eq:p_e2t} and~\eqref{eq:x0}, the total regret bound can then be calculated as
\begin{align}
 & R_k(T) \leq \Delta_k \biggl(  K\frac{\numberOfTaskTypes-1}{\numberOfTaskTypes} (\log(T) + 1) \nonumber\\
 & + 8\frac{\numberOfTaskTypes^5K^2}{\rho^{\numberOfTaskTypes^4+1}}\log(T)\biggl(\frac{e^{^{\frac{\Delta^2}{\numberOfTaskTypes\umax{}}}}}{1-\frac{\Delta^2}{\numberOfTaskTypes\umax{}}}  T^{1-\frac{\Delta^2}{\numberOfTaskTypes\umax{}}} + 6 \biggr) \biggr).
\end{align}
One can see that the leading order of the stable regret bound is a sublinear function which is given by
\begin{align}
    O \biggl( \Delta_k \frac{8\numberOfTaskTypes^5K^2 e^{^{\frac{\Delta^2}{\numberOfTaskTypes\umax{}}}}}{\rho^{\numberOfTaskTypes^4+1}(1-\frac{\Delta^2}{\numberOfTaskTypes\umax{}})} \log(T) T^{1-\frac{\Delta^2}{\numberOfTaskTypes\umax{}}} \biggr).
\end{align}
\end{appendices}

\vspace*{-0.6cm}
\section{Proof of Theorem 2}
\label{app:proofTheorem2}
Starting from \eqref{eq:probabilityNotStable}, we formulate
\begin{align}
    \sum_{t=1}^{T}\mathbb{P}(\boldsymbol{X}_t \notin X^{\mathrm{stable}}) & \leq \sum_{t=1}^{T}\mathbb{P}(E_{1,\timeindex}) + \sum_{t=1}^{T}\mathbb{P}(\overline{E}_{1,\timeindex})\mathbb{P}(E_{2,\timeindex}).
\end{align}
We use
\begin{align}
T \cdot \mathbb{P}(\boldsymbol{X}_T \notin X^{\mathrm{stable}}) \leq \sum_{t=1}^{T}\mathbb{P}(\boldsymbol{X}_t \notin X^{\mathrm{stable}})
\end{align}
as $\mathbb{P}(\boldsymbol{X}_t \notin X^{\mathrm{stable}})$ is monotonically decreasing in $t$.
Using \eqref{eq:p_e1t},~\eqref{eq:p_e2t} we can show that 
\begin{align}
   \mathbb{P}(\boldsymbol{X}_T \notin X^{\mathrm{stable}}) & \leq \frac{1}{T} \biggl(  K\frac{\numberOfTaskTypes-1}{\numberOfTaskTypes} (\log(T) + 1) \nonumber\\
 & + 8\frac{\numberOfTaskTypes^5K^2}{\rho^{\numberOfTaskTypes^4+1}}\log(T)\biggl(\frac{e^{^{\frac{\Delta^2}{\numberOfTaskTypes\umax{}}}}}{1-\frac{\Delta^2}{\numberOfTaskTypes\umax{}}}  T^{1-\frac{\Delta^2}{\numberOfTaskTypes\umax{}}} + 6 \biggr) \biggr)
\end{align}
One can see that the leading order of the probability $\mathbb{P}(\boldsymbol{X}_T \notin X^{\mathrm{stable}})$ is given by a function
\begin{align}
    O \biggl( \frac{8\numberOfTaskTypes^5K^2 e^{^{\frac{\Delta^2}{\numberOfTaskTypes\umax{}}}}}{\rho^{\numberOfTaskTypes^4+1}(1-\frac{\Delta^2}{\numberOfTaskTypes\umax{}})} \frac{\log(T)} {T^{\frac{\Delta^2}{\numberOfTaskTypes\umax{}}}} \biggr).
\end{align}
%
\end{document}